\documentclass[]{emulateica}

\pdfoutput=1

\usepackage%
[
pagebackref, 
colorlinks=true,
linkcolor=cyan,filecolor=cyan,
citecolor=cyan,
urlcolor=cyan,
pdftitle={Growth of Jupiter: Enhancement of core accretion},
pdfauthor={D'Angelo et al.},
pdfsubject={Growth of Jupiter: Enhancement of core accretion},
pdfcreator={KAZEditor},
pdfproducer={Kazzimiei},
pdfkeywords={Accretion; Jovian planets; Jupiter; Planetary formation; Planetesimals},
bookmarks=true,
breaklinks=true,
hypertexnames=false,
pdfstartview=FitV,
hyperfootnotes=true,
bookmarksopen=false,
pdfpagemode=None]{hyperref}
\usepackage{breakurl}

\usepackage{natbib}

\usepackage{graphicx}

\usepackage{amssymb}
\usepackage{textcase}

\makeatletter
\newcommand{\addspaces}[1]{%
  \@tfor\letter:=#1\do{%
    {\letter}\space
  }%
}
\makeatother

\topmatter{\href{http://dx.doi.org/10.1016/j.icarus.2014.06.029}{Icarus 241 (2014) 298-312}}
\shorttitle{\textit{\small{G. D'Angelo et al./Icarus 241 (2014) 298-312}}}
\shortauthors{\textit{\small{G. D'Angelo et al./Icarus 241 (2014) 298-312}}}

\begin{document}

\title{Growth of Jupiter: Enhancement of core accretion by
a voluminous low-mass envelope}


\author{Gennaro D'Angelo\altaffilmark{a,b,*,1},
        Stuart J. Weidenschilling\altaffilmark{c},
        Jack J. Lissauer\altaffilmark{a},
        Peter Bodenheimer\altaffilmark{d}
        }
\affiliation{~\\
$^{\mathrm{a}}$ \textit{%
                 NASA Ames Research Center, 
                 Space Science and Astrobiology Division, MS~245-3, 
                 Moffett Field, CA 94035, USA}\\
$^{\mathrm{b}}$ \textit{%
                SETI Institute, 189 Bernardo Ave., 
                 Mountain View, CA 94043, USA}\\
$^{\mathrm{c}}$ \textit{%
               Planetary Science Institute,
                 1700 East Fort Lowell Road, Suite 106, 
                 Tucson, AZ 85719-2395, USA}\\
$^{\mathrm{d}}$ \textit{%
                UCO/Lick Observatory, Department of Astronomy and
                 Astrophysics, University of California, Santa Cruz, 
                 CA 95064, USA}}
\altaffiltext{*}{Corresponding author at: 
                 NASA Ames Research Center, 
                 Space Science and Astrobiology Division, MS~245-3, 
                 Moffett Field, CA 94035, USA.
                 Fax: +1-650-604-6779.}
\altaffiltext{~}{\textit{E-mail address:} \href{mailto:gennaro.dangelo@nasa.gov}{gennaro.dangelo@nasa.gov} (G.~D'Angelo).}
\altaffiltext{1}{Address: Los Alamos National Laboratory, Los Alamos, NM 87545, USA.}
%



\begin{abstract}
\vspace*{-3mm}%
\hspace*{-16mm}\rule{\textwidth}{0.3mm}\\

\hspace*{-16mm}\begin{minipage}[t]{0.25\textwidth}
\addspaces{ARTICLE}~~\addspaces{INFO}\\
\rule{\textwidth}{0.3mm}\\
\begin{small}
\textit{Article history:}\\
Received 26 August 2013\\
Revised 27 June 2014\\
Accepted 27 June 2014\\
Available online 10 July 2014\\
\end{small}
\rule{\textwidth}{0.3mm}\\
\begin{small}
\textit{Keywords:}\\
Accretion\\
Jovian planets\\
Jupiter\\
Planetary formation\\
Planetesimals
 
\begin{flushleft}
\footnotesize{%
Article in press.
This is an unofficial preprint prepared by the authors 
from the accepted manuscript.
             }
\end{flushleft}
\end{small}
\end{minipage}\hspace*{0.05\textwidth}%
\begin{minipage}[t]{0.693\textwidth}
\addspaces{ABSTRACT}\\
\rule{\textwidth}{0.3mm}\\
\begin{small}
We present calculations of the early stages of the formation of Jupiter 
via core nucleated accretion and gas capture. 
The core begins as a seed body of about $350$ kilometers in radius 
and orbits in a swarm of planetesimals whose initial radii range from 
$15$ meters to $50$ kilometers.
The evolution of the swarm accounts for growth and fragmentation, 
viscous and gravitational stirring, and for drag-assisted migration and 
velocity damping. 
During this evolution, less than $9$\% of the mass is in planetesimals smaller
than $1$ kilometer in radius; $\lesssim 25$\% is in planetesimals with 
radii between $1$ and $10$ kilometers; and $\lesssim 7$\% is in bodies
with radii larger than $100$ kilometers.
Gas capture by the core substantially enhances the size-dependent 
cross-section of the planet for accretion of planetesimals.
The calculation of dust opacity in the planet's envelope accounts 
for coagulation and sedimentation of dust particles released 
as planetesimals are ablated. The calculation is carried out at an
orbital semi-major axis of $5.2\,\mathrm{AU}$ and the initial solids' surface 
density is $10\,\mathrm{g\,cm^{-2}}$ at that distance.
The results give a core mass of nearly $7.3$ Earth masses ($\mathrm{M}_{\oplus}$) 
and an envelope mass of $\approx 0.15\,\mathrm{M}_{\oplus}$ after about 
$4\times 10^{5}$ years, at which point the envelope growth rate surpasses 
that of the core. 
The same calculation without the envelope yields a core of only about 
$4.4\,\mathrm{M}_{\oplus}$.\\[-3mm]
\begin{flushright}\copyright\ 2014 Published by Elsevier Inc.\end{flushright}
\end{small}
\end{minipage}\\[4mm]
\hspace*{-13mm}\rule{\textwidth}{0.3mm}\\
\end{abstract}


\section{Introduction}
\label{sec:introduction}
\defcitealias{bodenheimer1986}{BP86}
\defcitealias{bodenheimer2000b}{BHL00}
\defcitealias{hubickyj2005}{HBL05}
\defcitealias{pollack1996}{PHBLPG96}
\defcitealias{naor2010}{MBPL10}
\setcounter{footnote}{1}
The formation of Jupiter is a key element in the classical problem of the origin 
of the Solar System. Detailed studies of the formation of this planet by 
core-nucleated accretion have been carried out for decades
\citep{safronov1972,perri1974,mizuno1980,bodenheimer1986,pollack1996}. 
The  latter work studied what is now considered to be a standard case: the
formation of  Jupiter, in a fixed orbit at $5.2\,\mathrm{AU}$, in a disk with solid 
surface density $\sigma_{Z}=10\,\mathrm{g\,cm^{-2}}$, about three times as 
high as that in the minimum-mass solar nebula \citep{weidenschilling1977a,hayashi1981}.
\citeauthor{pollack1996}'s basic conclusion was that the formation time can range from
$1.25$ to $8\,\mathrm{Myr}$, depending on physical assumptions made in 
the computations.
The heavy-element core masses fell in the range $12$--$20$ Earth masses
($\mathrm{M}_{\oplus}$).

Jupiter's growth involves numerous elements of physics over a wide range of mass 
and length scales. The initial steps in the process involve the buildup of
planetesimals from an initial distribution of sub-micron-size dust grains
\citep[e.g.,][]{chiang2010}. The work described in the present paper
starts at a somewhat later stage, when a swarm of planetesimals, with radii
ranging from several meters to tens of kilometers, has formed, along with a nascent 
planetary embryo of somewhat larger size. The embryo (composed almost entirely 
of elements heavier than hydrogen and helium) builds up by accretion of
planetesimals and becomes the planetary core. When its mass reaches 
$\sim 0.1\,\mathrm{M}_{\oplus}$, and when the escape speed from its surface 
exceeds the thermal speed of molecules in the surrounding gas disk, 
it begins to capture a small amount of gas.  
This gas is assumed to have nearly solar composition. 
However, the accretion rate of solids onto the core ($\dot{M}_{c}$) greatly exceeds, 
for some time, the accretion rate of gas ($\dot{M}_{e}$, composed primarily 
of molecular hydrogen and helium). 
Once the core has accreted most of the planetesimals within its gravitational reach,
$\dot{M}_{c}$ slows down significantly and $\dot{M}_{e}$, which is increasing, 
begins to exceed it. Thereafter, during a phase characterized by slow gas 
accretion, $M_{c}$ continues to grow, but less rapidly than $M_{e}$, until 
the \textit{crossover mass} is reached ($M_{c}= M_{e}$).
We denote by ``Phase~1'' the time up to the point where 
$\dot{M}_{e} =\dot{M}_{c}$, and by ``Phase~2'' the time from there up to crossover.
At or before crossover,  the rapid gas accretion phase begins 
($\dot{M}_{e} \gg \dot{M}_{c}$). The gas accretion rate is at
first limited by the rate at which the envelope can contract and release
energy, governed primarily by the opacity due to dust and gas. 
The contraction rate increases and soon reaches the 
point where the accretion rate required by the 
contraction exceeds the rate at which the disk can
provide gas. The \textit{disk-limited} accretion phase begins, 
with the gas accretion rate depending on several factors:
planet mass, planet orbital radius, 
disk gaseous density, disk kinematic viscosity, 
and disk scale height. The total planet mass ($M_{p}$)
at which the disk-limited phase starts is typically several
tens of Earth masses \citep{lissauer2009}. 
Disk-limited accretion rates are determined by three-dimensional 
hydrodynamics simulations of a planet embedded in a disk 
\citep[and references therein]{lissauer2009,bodenheimer2013}.
The final mass of the planet is determined by a combination 
of gap opening, which drastically reduces accretion,
and the dissipation of the nebular gas.

In the case of Jupiter, there are several observational constraints that
must be satisfied by any formation model. Protoplanetary disk lifetimes
have a median value of $\sim 3\,\mathrm{Myr}$ and a maximum of 
$\sim 10\,\mathrm{Myr}$ \citep{hillenbrand2008,roberge2011}. 
The mass ($\mathrm{M}_{\mathrm{J}} = 1.898 \times 10^{30}\,\mathrm{g}$), 
equatorial radius ($\mathrm{R}_{\mathrm{J}} = 7.15 \times 10^{9}\,\mathrm{cm}$) 
and gravitational moments $J_{2}$, $J_{4}$, $J_{6}$ are measured, constraining 
the mean density and density distribution.  The Galileo probe measured 
the abundances of a number of elements in Jupiter's outer layers 
\citep{young1996,owen1999,young2003} 
and determined that they were in the range of $2$--$4$ times solar.
Derived core masses from models of the interior of Jupiter vary considerably 
depending on the equation of state and assumed composition layering. 
\citet{militzer2008} obtain  $M_{c}=16 \pm 2\,\mathrm{M}_{\oplus}$;
\citet{nettelmann2012} find $M_{c}=0$--$8\,\mathrm{M}_{\oplus}$;
\citet{saumon2004} find $M_{c}=0$--$11\,\mathrm{M}_{\oplus}$.
The Nettelmann three-layer models are consistent with the abundances 
measured by Galileo in the atmosphere, and have total heavy-element 
masses in the range $28$--$32\,\mathrm{M}_{\oplus}$, 
about six times solar. 
However, this quantity is not well constrained.  Estimates range from $0$ to 
$18\,\mathrm{M}_{\oplus}$ for the core mass $M_{c}$, and  from $15$ to 
$40\,\mathrm{M}_{\oplus}$ for the total heavy-element mass $M_{Z}$
\citep{fortney2010}. 
In this paper, we do not distinguish between the two masses $M_{c}$ and $M_{Z}$.

Several major improvements in the physical basis of the computations have
been made since the work of \citet{pollack1996}. \citet{alibert2005a}
and \citet{mordasini2012} include the effects of disk evolution and
planetary orbital migration on the formation process. \citet{lissauer2009}
and \citet{bodenheimer2013} use gas accretion rates for the disk-limited
phase based on three-dimensional hydrodynamic simulations of disk
flow around an embedded planet. 
\citet[hereafter \citetalias{naor2010}]{naor2010} calculate the
dust opacity in the envelope of the forming planet self-consistently
by including the effects of dust settling and coagulation. 
The results of \citetalias{naor2010} 
show a significant increase in gas accretion rate prior to the
disk-limited phase, due to the reduced opacity (and hence faster cooling) 
in the outer parts of the planet's envelope. 
\citet{inaba2003b} include a statistical treatment of
the planetesimal accretion rate onto the core, an improvement over the
more approximate treatment of this rate by \citet{pollack1996}. The time
to form the initial core is crucial in determining the formation time
for the entire planet. They  include the enhancement in the capture
cross-section for planetesimals as a result of the presence of the gaseous
envelope, as well as collisional fragmentation of planetesimals and a
range of planetesimal sizes.
\citet{inaba2003b} came to the basic conclusion that a very high surface
density of solid material in the initial disk, $\sigma_{Z}=25\,\mathrm{g\,cm^{-2}}$ 
 at $5.2\,\mathrm{AU}$, was required to build a core of $21\,\mathrm{M}_{\oplus}$ 
within the lifetime of the protoplanetary disk. This core mass  is above the
values required to initiate rapid gas accretion (which they did not calculate)
but their  $\sigma_{Z}$ implies a disk mass well above typical observed values.
The above result was obtained with assumed interstellar grain opacities
in the planet's atmosphere. If these opacities are reduced by a factor $100$, 
they find that with a smaller value of the surface density,
$\sigma_{Z}=12.5\,\mathrm{g\,cm^{-2}}$, a core of $7\,\mathrm{M}_{\oplus}$
can form at $5.2\,\mathrm{AU}$ in $5\,\mathrm{Myr}$, 
probably sufficient to collect gas.

The results of \citet{inaba2003b} imply that core formation times at
$5.2\,\mathrm{AU}$  are actually longer than those calculated by, for example, 
\citet{pollack1996} with an  assumed $\sigma_{Z}=10\,\mathrm{g\,cm^{-2}}$.
\citeauthor{inaba2003b} also confirm the findings of 
\citeauthor{pollack1996} that the grain opacity is an important quantity during
initial core formation. 
If the opacity is reduced, the envelope density must
increase to maintain hydrostatic and thermal equilibrium during Phase~1, 
at a given core accretion rate; thus, more gas flows into the envelope and 
the capture cross-section for planetesimals is enhanced. 

In this paper we consider the initial core formation, Phase~1,
at $5.2\,\mathrm{AU}$ with $\sigma_{Z}=10\,\mathrm{g\,cm^{-2}}$, 
a reasonable value for an initial solar nebula \citep{weidenschilling2005}. 
The main question to be answered is whether or not a core of $5$--$10\,\mathrm{M}_{\oplus}$  
can be formed on a timescale of less than $1$--$2\,\mathrm{Myr}$. 
An initial core (at the end of Phase~1) in that mass range is likely to accrete gas 
and build up to a Jupiter mass in less than a typical disk lifetime \citepalias{naor2010}. 
To answer that question we combine two state-of-the-art codes, 
one for the statistical treatment of planetesimal accretion, and the other
for the calculation of the structure, evolution, and capture cross-sections
of the planet's gaseous envelope. The grain opacity in the envelope includes the
effects of coagulation and settling.  The two codes interact, in that the 
planetesimal code provides $\dot{M}_{\mathrm{c}}$ while the envelope code
provides $M_{\mathrm{e}}$ and the capture radius for each planetesimal
size in the assumed range. The details of the codes are presented
in Section~\ref{sec:NP}.  
Previous relevant studies and some of their findings are briefly reviewed
in Section~\ref{sec:compa}. Our results are discussed in Section~\ref{sec:results}, 
and compared with parallel calculations (\textit{i}) without the presence of the envelope, 
and (\textit{ii}) with the envelope but with the core accretion rates used in previous works including
\citet{pollack1996} and \citetalias{naor2010}. The conclusions are presented in 
Section~\ref{sec:summary}.

\section{Numerical procedures}
\label{sec:NP}

In this section, we outline the main numerical procedures
that are employed in our models of Jupiter's core accumulation 
and envelope formation.

\subsection{The planetesimal accretion code}
\label{sec:PAC}

A detailed discussion of the solid-body accretion code is presented in 
\citet[see also \citeauthor{weidenschilling2011}, \citeyear{weidenschilling2011}]{weidenschilling1997}. Here we present a general summary of the method and 
the procedure as used in the simulations. 
The accretion code computes collisional and
gravitational interactions within a swarm of planetesimals extending 
over a wide range of heliocentric distances. The swarm is divided into a 
number of radial zones, each corresponding to a narrow range of semi-major 
axes. Within each of these zones, the size-frequency distribution is
represented by the number of bodies in each of a series of logarithmic 
diameter bins, chosen so that the mean mass differs by a factor of two
between adjacent bins. Each bin has mean values of orbital eccentricity and
inclination, with an assumed range about the mean. During a time-step,
all zone/bin combinations that have crossing orbits are identified.
Impact probabilities are computed, based on the relative velocities,
gravitationally enhanced collisional cross-sections, and the fractional
overlap of orbits. Based on these probabilities, the locations of a set 
of typical collisions are selected stochastically, and the corresponding
impact velocities are evaluated. The outcome of an impact depends on
the velocity, the mass ratio of projectile and target, and their 
size-dependent impact strength, which includes gravitational binding
energy. Either or both bodies may be eroded or shattered. Fragments
are assumed to have a power-law size distribution; for shattering events
the slope of this distribution depends on the specific energy. A fraction ($1$\%)
of the impact energy in the center of mass frame is partitioned into
kinetic energy of the fragments. Depending on the escape velocity of the
colliding bodies, some fraction of the mass may escape, or it all may be accreted.
The merged body is assumed to have the velocity of the center of mass,
while escaping fragments are assigned a mean velocity with a randomly
chosen direction relative to that vector. New orbits are computed from 
these velocity components, and the appropriate amounts of mass are 
added to the corresponding bins and zones. Thus, collisions can transfer
mass between radial zones, in addition to causing the size distribution
to evolve with time. 

Eccentricities and inclinations evolve separately due to viscous stirring
and dynamical friction. The rates for velocity evolution due to interactions 
among the bodies in the swarm are modeled by the formulae of \citet{wetherill1989}. 
Collisions, modeled as described above, have a net damping
effect on velocities. As nebular gas is assumed to be present during Jupiter's
formation, gas drag also acts to damp eccentricities and inclinations 
\citep{adachi1976}. As the accretion code assumes Keplerian dynamics for the
planetesimals, it is not realistic to track the fates of bodies smaller
than $\sim 10$ meters in size, for which gas drag causes rapid orbital
evolution. 
Such bodies would presumably be lost by spiraling inward toward the Sun
on short timescales.
In fact, denoting with $a$, $\Omega$, and $R$ the initial semi-major axis,
orbital frequency, and radius of the body, and assuming a drag coefficient of
order unity, the timescale $\tau_{\mathrm{drag}}$ for orbital decay due 
to gas drag is
\begin{equation}
\Omega \tau_{\mathrm{drag}}\sim \frac{16}{3}
                                             \left(\frac{\rho_{s}}{\rho_{\mathrm{neb}}}\right)
                                             \left(\frac{R}{a}\right)
                                             \left(\frac{a}{H}\right)^{4},
\label{eq:tau_drag}
\end{equation}
where $\rho_{s}$ and $\rho_{\mathrm{neb}}$ indicate the density of the body
and the nebular gas, respectively. The ratio $H/a$ is the local aspect ratio 
of the nebula (the relative pressure scale height), and quantifies the deviation 
from Keplerian rotation 
($\Omega_{\mathrm{K}}$) of the gas. Typically, $\Omega$ is such that
$(\Omega_{\mathrm{K}}-\Omega)/\Omega_{\mathrm{K}}\sim (H/a)^{2}$.
For $10\,\mathrm{m}$-size bodies,
the conditions used here (see Section~\ref{sec:EPS}) result in a variation
$\gtrsim 10$\% in semi-major axis over a few tens of orbital periods.
We assume that all fragments smaller than $15\,\mathrm{m}$ are lost; 
in our simulations this loss is typically $< 10$\% of the total mass of the swarm.

The assumption that the majority of these small fragments are lost, and not
accreted by the dominant body, or core, can be proven by showing that
the time to cross (in the radial direction) the core's libration region of half-width 
$w$, $\tau_{\mathrm{cross}}\sim\tau_{\mathrm{drag}}(2w/a)$,
is much shorter than the libration timescale of a fragment,
$\tau_{\mathrm{lib}}=8\pi a/(3\Omega w$). 
Using Equation~(\ref{eq:tau_drag}), the ratio of the two timescales becomes
\begin{equation}
\frac{\tau_{\mathrm{cross}}}{\tau_{\mathrm{lib}}}\sim
                                             \left(\frac{\rho_{s}}{\rho_{\mathrm{neb}}}\right)
                                             \left(\frac{R}{a}\right)
                                             \left(\frac{a}{H}\right)^{4}
                                             \left(\frac{w}{a}\right)^{2}.
\label{eq:tau_rat}
\end{equation}
When $w$ is about equal to the Hill radius, 
$R_{\mathrm{H}}=a\left[M_{p}/(3 M_{\odot})\right]^{1/3}$, the ratio
in Equation~(\ref{eq:tau_rat}) is $\sim 0.01$ for a $10\,\mathrm{m}$-size fragment
and a $5$ Earth-mass core, and its value is even less for smaller fragment 
size and/or core mass.
More detailed calculations leading to the same conclusion are presented by 
\citet{kary1993,kary1995}. 

The evolution of the planetesimal swarm is dominated by its interactions 
with the growing Jovian core, which becomes much larger than the remaining
bodies. Due to conservation of the Jacobi parameter in the restricted 3-body
problem, a small body whose orbit crosses that of a single massive body is 
stirred less effectively than one that can encounter more than one massive
body. In principle, the Jacobi parameter allows close encounters for orbits
that are initially separated in semi-major axis by up to $2 \sqrt{3}\,R_{\mathrm{H}}$. 
However, only a limited range of separations between about $1.8$ and 
$2.4\,R_{\mathrm{H}}$ allow a planetesimal with an initially circular orbit to enter 
the core's Hill sphere in a single encounter \citep{nishida1983}. For orbits in 
this range we use the formalism of \citet{greenzweig1990,greenzweig1992} for 
gravitational stirring. For more distant orbits, the stirring rate is computed 
according to \citet{weidenschilling1989}. Smaller separations are in horseshoe or 
Trojan-type orbits, and do not approach the core. Such bodies comprise a small 
fraction of the swarm's mass, and do not affect the core's evolution (they 
should not be identified with the present Trojan asteroids, which were probably 
captured subsequent to Jupiter's formation). 

Planetesimals are brought to the vicinity of the core primarily by Keplerian
shear, rather than by their random velocities. They encounter the core at a rate 
inversely proportional to their synodic period, or proportional to the 
difference in their semi-major axes. During an encounter, the velocity impulse 
imparted by the core is mostly in the plane of its orbit, so the stirring rate 
for inclinations is much less than for eccentricities. This effect tends to 
keep the planetesimal swarm relatively flat. Stirring within a swarm of
comparably-sized bodies typically produces inclinations with magnitude
about half that of eccentricities, but in the region within a few Hill radii of
the core's semi-major axis, eccentricities are about an order of magnitude 
larger than inclinations. Planetesimals with nonzero eccentricities are able 
to enter the core's Hill sphere if their perihelia or aphelia are within
$2.4\,R_{\mathrm{H}}$ of the core's orbit. We compute the probability of collision with 
the core according to \citet{greenberg1991}.  Appropriate expressions are 
used, depending on whether the thickness of the swarm (semi-major axis times 
inclination) is larger or smaller than the Hill radius and/or the 
gravitationally enhanced collisional radius.

\subsection{Evolution of the planetesimal swarm}
\label{sec:EPS}

We start the growth of Jupiter's core from a ``seed'' body embedded in a 
swarm of smaller planetesimals at Jupiter's present distance, $a=5.2\,\mathrm{AU}$.
The radial grid for the planetesimal evolution calculation extends from $4.7$ 
to $5.75\,\mathrm{AU}$.
We impose the condition that the bodies in the outermost zone are not depleted 
due to gas drag; i.e., that the bodies in that zone are effectively replaced by bodies 
from farther out with the same size distribution.
The nebula's surface density varies inversely with heliocentric
distance; the value for the planetesimal swarm is 
$\sigma_{Z}=10\,(5.2\,\mathrm{AU}/a)\,\mathrm{g\,cm^{-2}}$, 
giving an isolation mass \citep{lissauer1987}
\begin{equation}
M_{\mathrm{iso}}\approx 0.0026 \left(\frac{a}{\mathrm{AU}}\right)^{3}%
\left(\frac{\sigma_{Z}}{\mathrm{g\,cm^{-2}}}\right)^{3/2}\,\mathrm{M}_{\oplus}
\label{eq:Miso}
\end{equation}
of about $11\,\mathrm{M}_{\oplus}$, assuming a ``feeding zone'' of full width
equal to $8\,R_{\mathrm{H}}$ \citep[see][]{lissauer1993b,kary1994}, where
the Hill radius is defined after Equation~(\ref{eq:tau_rat}). 
The surface density of the gas at the seed's orbit is 
$\sigma_{XY} = 1000\,\mathrm{g\,cm^{-2}}$, with density 
$3.3\times10^{-9}\,\mathrm{g\,cm^{-3}}$. 
The radial pressure gradient causes the gas to revolve about the proto-sun 
at $\sim 25\,\mathrm{m\,s^{-1}}$ less than the local Keplerian velocity. 

The initial swarm is assumed to consist 
of planetesimals with a power law size distribution from $15\,\mathrm{m}$ to 
$50\,\mathrm{km}$ in radius; the slope is the collisional equilibrium value of 
$-11/6$ (incremental mass), which places most of the mass in the largest 
bodies, tens of kilometers in radius or larger. 
Their bulk density is taken to be constant
and equal to $1.4\,\mathrm{g\,cm^{-3}}$,
corresponding to a mixed rock/ice composition, with the mass of the
largest planetesimals being $7.7\times 10^{20}\,\mathrm{g}$.
The planetesimal fragmentation model takes into account both 
a size-dependent material strength and the gravitational binding energy
\citep{davis1994}. 
For the smallest bodies, the material strength $S$ decreases with increasing 
radius $R$ due to the presence of defects in the solid material; we assume 
that $S(R)$ is proportional to $R^{-0.24}$ in this regime. At large sizes, 
self-compression increases strength, with $S(R)$ proportional to $R^{1.89}$.
The minimum strength is $S = 1.9\times 10^{5}\,\mathrm{erg\,g^{-1}}$ at 
$R = 5\,\mathrm{km}$. Impacts result in cratering or shattering of planetesimals.
The impact strength essentially governs the size
distribution of fragments. The amount of mass escaping is set by the ratio of the 
kinetic energy of the fragments to the impact energy in the center-of-mass 
frame. We assume a ratio of $1$\%, i.e, highly dissipative collisions. 
As most of the mass of the swarm is in planetesimals tens of kilometers 
in size, only a small fraction of the swarm mass is lost by collisional grinding 
down to radii $\lesssim 10\,\mathrm{m}$. The results of the simulations are 
not sensitive to the strength of the planetesimals. 
All planetesimals striking the core are accreted. 

The initial orbital eccentricity of the planetesimals is $\sim 1.7\times 10^{-3}$, 
with the initial inclination half that value;  these values give initial random velocities 
comparable to the escape velocity of the largest planetesimals. 
The initial values are not critical, provided they are low enough so that the
initial accretion by the seed body is dominated by Keplerian shear, which allows 
the core to outgrow the other bodies. As runaway growth occurs, velocities near 
the core's orbit are dominated by its perturbations, not by those of the bodies in the 
swarm. The core has a much lower eccentricity, $\sim 10^{-4}$, throughout the simulation.

The seed body has an initial mass
of $5.7\times 10^{23}\,\mathrm{g}$ ($10^{-4}\,\mathrm{M}_{\oplus}$) 
and density $3.2\,\mathrm{g\,cm^{-3}}$, for a radius of 
$\sim 350\,\mathrm{km}$; the density of the growing planetary core
remains constant throughout the simulation presented herein, with
compression compensating for the accretion of lower density planetesimals. 
This roughly Ceres-sized body is sufficiently large compared with the
neighboring planetesimals to initiate runaway growth. The initial
conditions ensure that the growth of the core occurs in the regime
dominated by Keplerian shear, with the swarm's thickness smaller than 
the size of its Hill sphere. This regime allows the most rapid 
``monarchical'' growth \citep{weidenschilling2005}. Although other bodies in the
simulation are allowed to grow, they are unable to overtake the core
before its own perturbations stir eccentricities in the surrounding region
and prevent runaway growth of potential competitors. In a typical simulation,
the largest bodies other than the core reach sizes of a few hundred km
in radius.
We do not speculate on the origin of the seed body, which is more massive 
than its neighbors by three orders of magnitude. However, we note that such 
a body is more plausible than the Mars- or Earth-sized initial bodies assumed
in many previous studies
(e.g., \citeauthor{pollack1996},  \citeyear{pollack1996}; 
\citeauthor{dodson2008},  \citeyear{dodson2008};
\citetalias{naor2010}).

We might expect the rapid growth of the core under these conditions
would continue until it reached the isolation mass, $\sim 10\,\mathrm{M}_{\oplus}$.
However, that limit is based on the idealized restricted
three-body problem, in which there are no interactions among planetesimals
and no non-gravitational forces. We find that inclusion of mutual
collisions among the planetesimals and nebular gas drag affect the later
stages of core growth, and may prevent it from reaching the theoretical isolation 
mass in Equation~(\ref{eq:Miso}). 
If a planetesimal experiences a synodic encounter with the core that does not
result in a collision, its eccentricity is typically increased. Conservation
of the Jacobi parameter implies that its semi-major axis is also changed
in such a manner as to increase its mean distance from the core; i.e, if
its orbit is inside (outside) that of the core, its semi-major axis decreases
\citep[increases; see][]{nishida1983}. 
This effect occurs for synodic encounters at all distances, not 
just for planetesimals in the core's feeding zone; however, it is strongest 
near the core's orbit. In the absence of dissipation, the increase in the 
planetesimal's eccentricity allows a planetesimal capable of a close encounter
with the core to make additional close approaches in future encounters. However, 
inclusion of damping, by collisions or gas drag, may reduce its eccentricity
before the next synodic encounter. The net result is to push planetesimals away
from the core 
\citep{greenberg1983}. 
This ``shepherding'' effect, in combination with the depletion of mass due
to accretion, produces a gap in the swarm around the orbit of the growing core
\citep{tanaka1997}.
Note that this gap is only in the swarm of solid planetesimals; the
planet does not typically produce a gap in the nebular gas until it approaches 
a mass of a few to several times $10\,\mathrm{M}_{\oplus}$, depending on
nebular parameters, see \citet{gennaro2011,lubow2011}. 

Planetesimals pushed away by the core's perturbations pile up 
in the regions adjacent to the gap, increasing the local surface density of
solids. This effect is partially offset by the secular decay of semi-major 
axes due to gas drag, on a timescale with order of magnitude given 
by Equation~(\ref{eq:tau_drag}),
as the radial pressure gradient causes non-Keplerian 
rotation of the nebula gas. Depending on their sizes and the nebular parameters, 
small bodies migrating inward may pile up near the outer edge of the gap, 
or be fed to the core from that region. On the sunward side, the core's 
perturbations and drag act in the same direction, softening somewhat 
the edge of the gap.
\begin{figure*}
\centering%
 \resizebox{.80\hsize}{!}{\includegraphics[angle=00,clip]{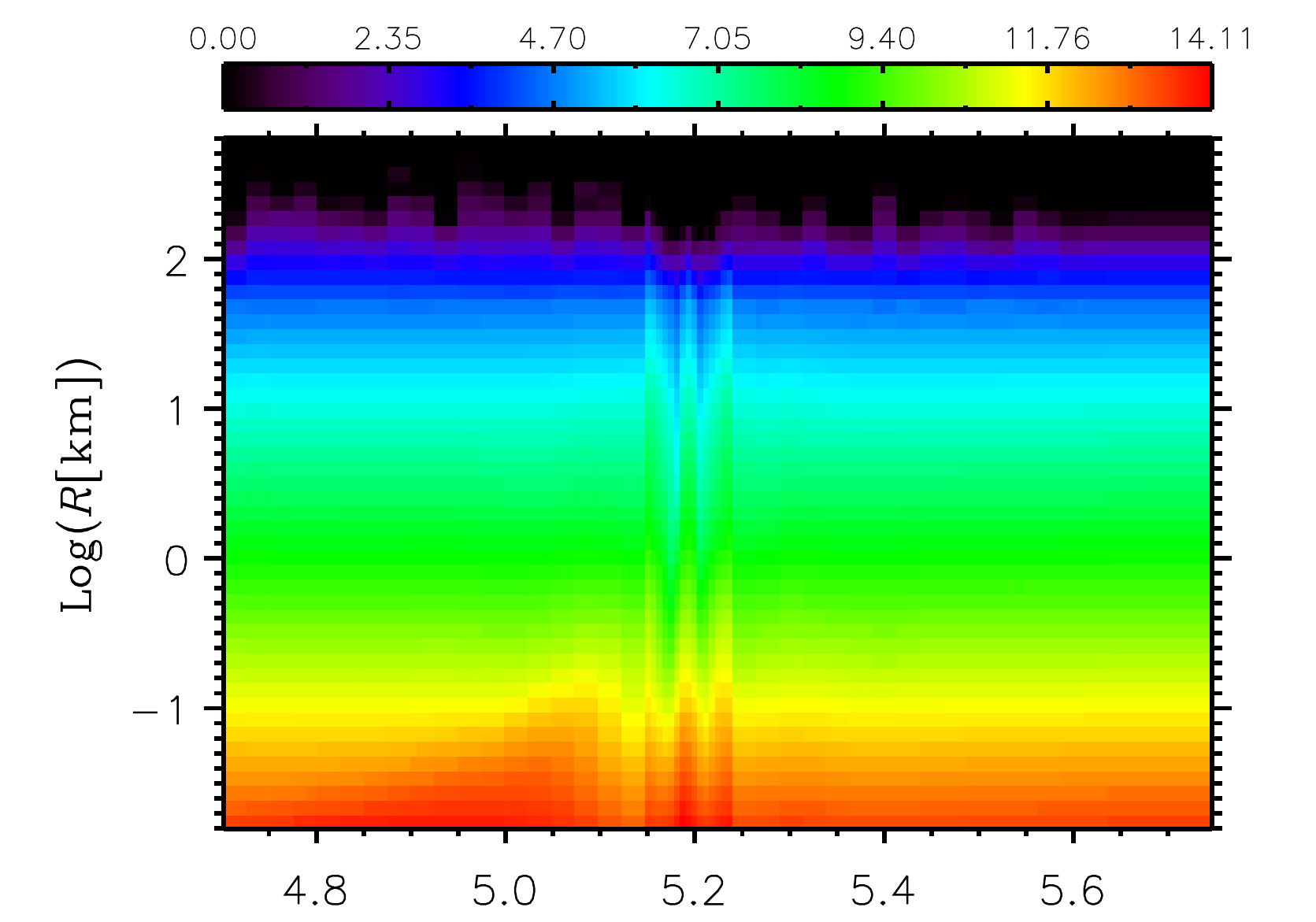}%
                                  \includegraphics[angle=00,clip]{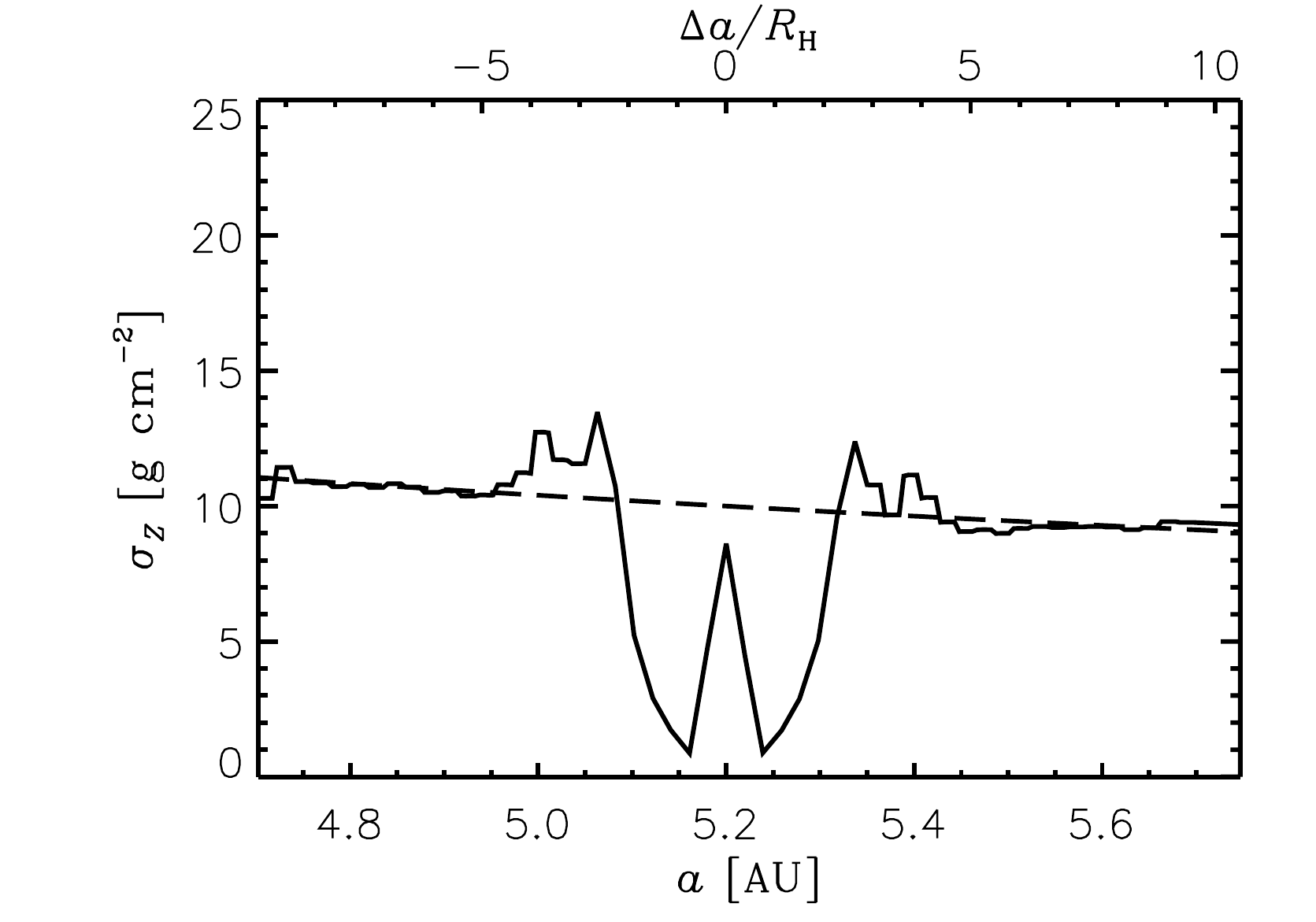}}
 \resizebox{.80\hsize}{!}{\includegraphics[angle=00,clip]{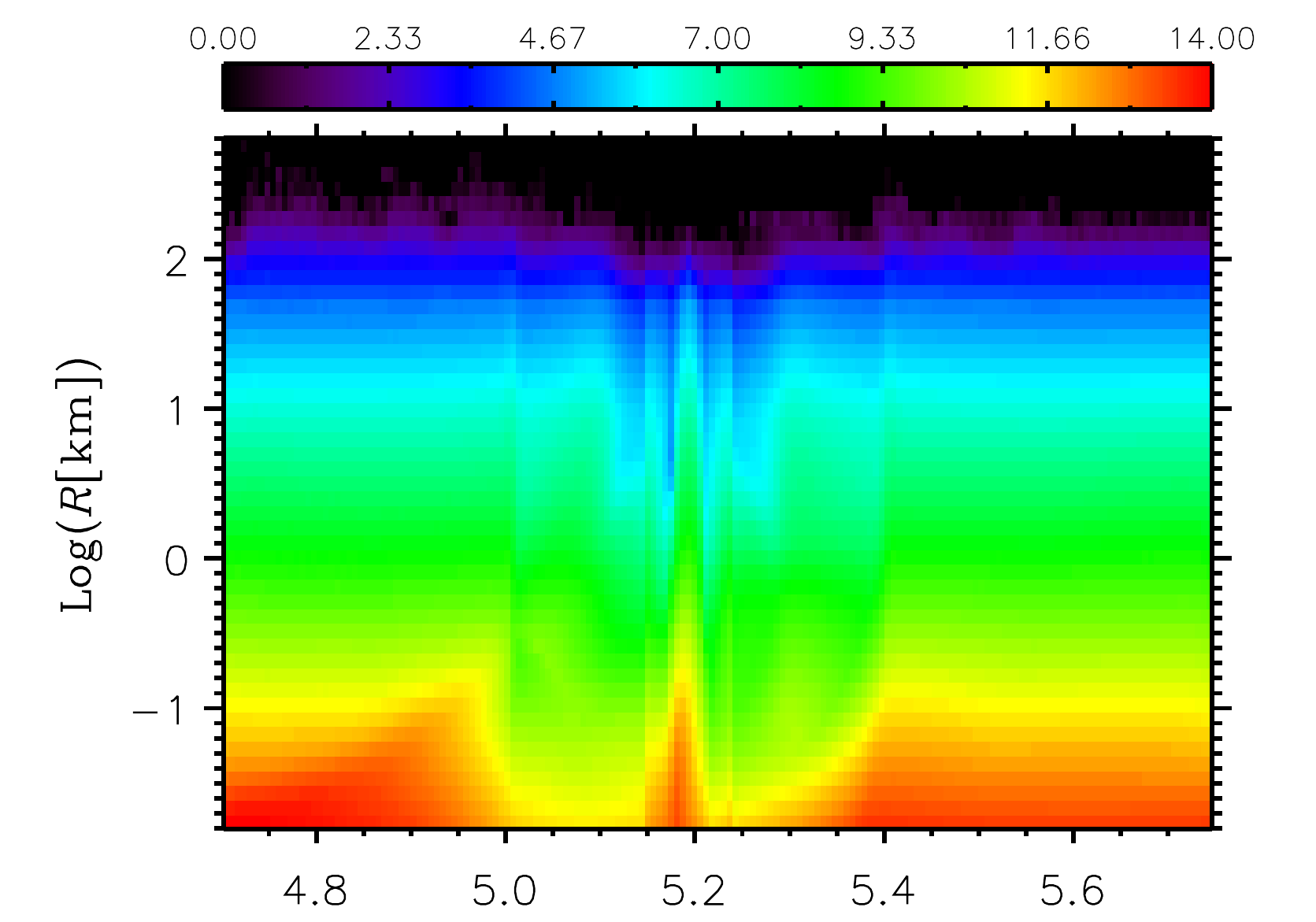}%
                                  \includegraphics[angle=00,clip]{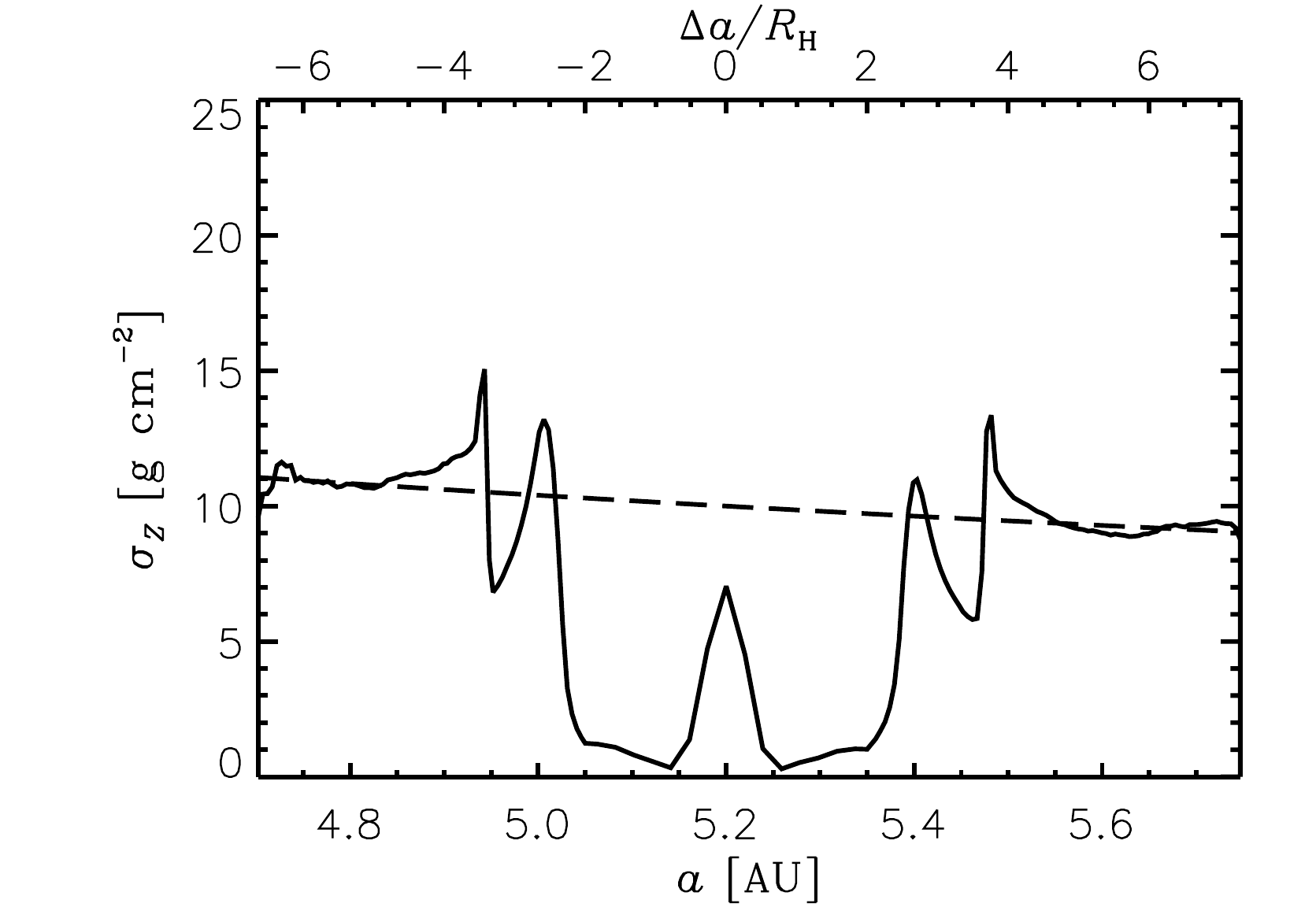}}
 \resizebox{.80\hsize}{!}{\includegraphics[angle=00,clip]{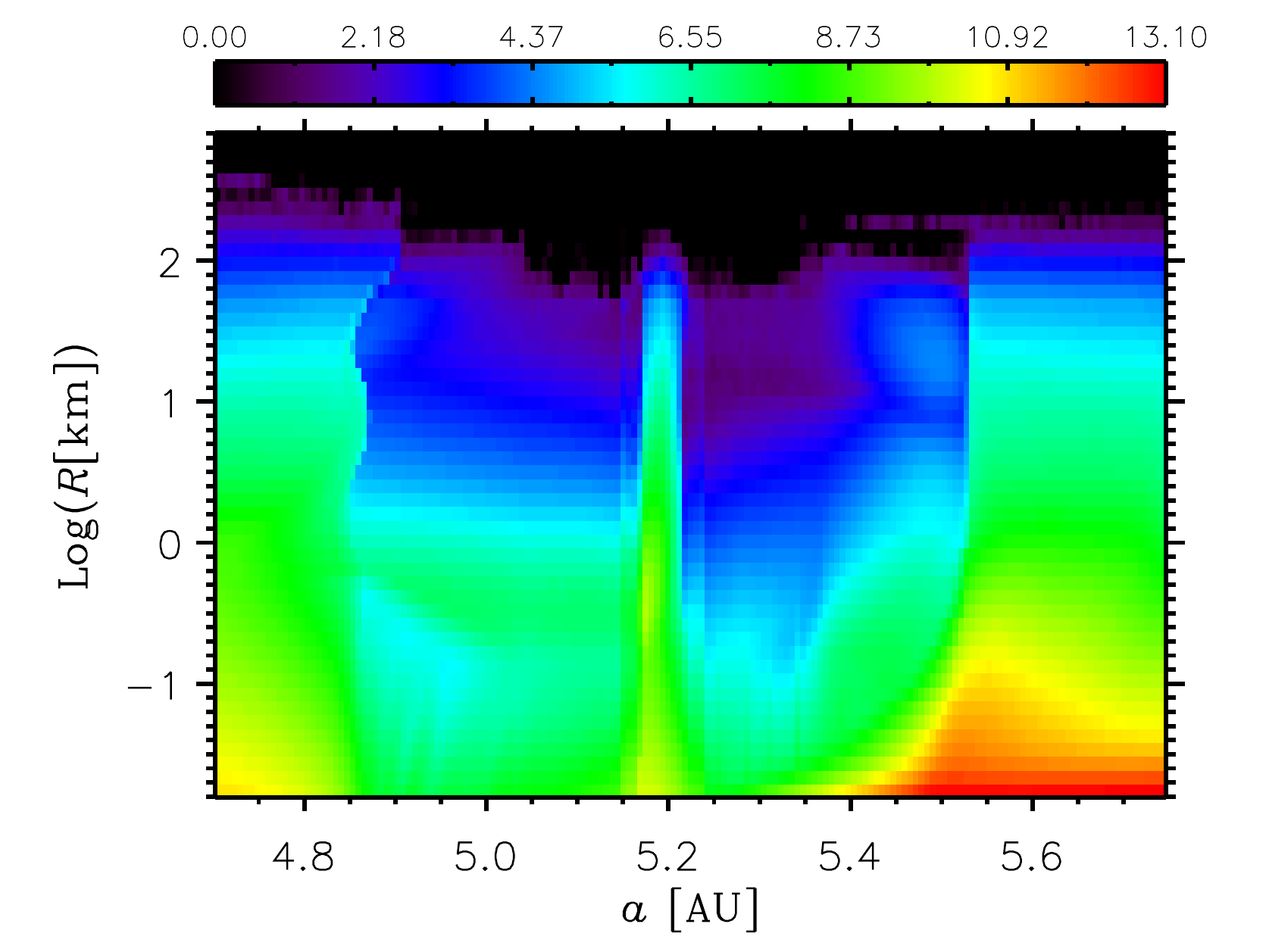}%
                                  \includegraphics[angle=00,clip]{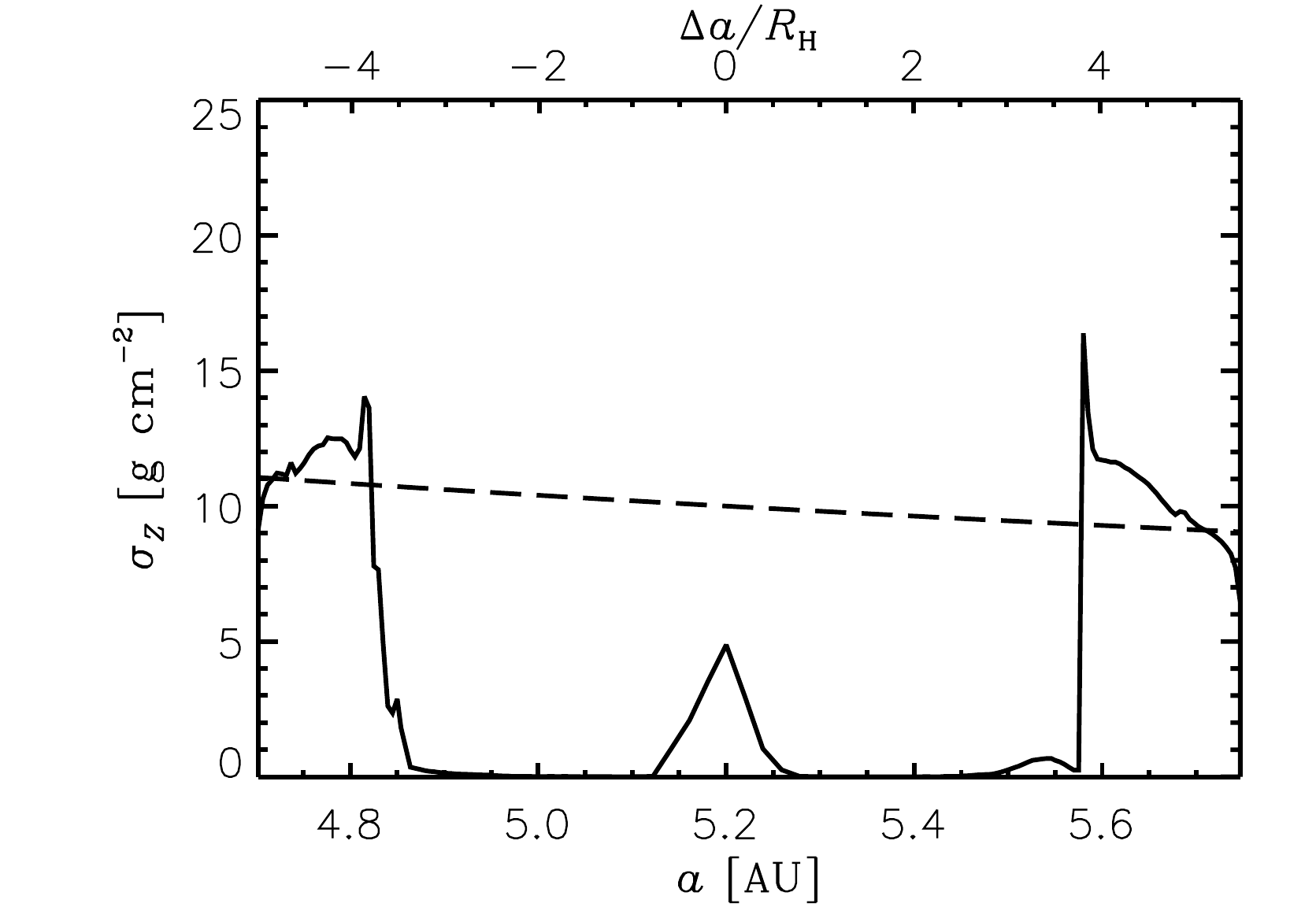}}
  \caption{%
               Left. Planetesimal distributions as a function of semi-major axis and radius.
               Each plot shows the logarithm (in base $10$) of the number of bodies per 
               radial zone and size bin. 
               From top to bottom, the distributions are plotted when $M_{c}=1$, $3$, and 
               $7\,\mathrm{M}_{\oplus}$. 
               Right. Cumulative surface density per radial zone (solid lines), 
               computed as the total planetesimal mass divided by the surface area 
               of the zone. The dashed line represents the initial
               surface density of solids, which scales as $1/a$. The top axis of
               each panel gives the distance from the planet in units of the planet Hill radius.
               These plots clearly
               show the formation and widening of the a gap in the swarm around
               the orbit as the planet grows.
               }
  \label{fig:maps}
\end{figure*}
These effects can be seen in Figure~\ref{fig:maps}, which illustrates the
number of planetesimals for each zone in semi-major axis and for each
size bin (left) and the surface density of solids versus semi-major axis 
(solid lines in the right panels).
There are modest asymmetries in the planetesimal distribution
with respect to the core's orbit (see, e.g., bottom left panel).
The formation of a gap in the swarm about the orbit of the core, and its 
increasing width as the core grows, is clearly visible in the right panels,
as is the local increase in surface density in the regions near the 
gap edges. The half-width of this gap is about $4\,R_{\mathrm{H}}$
when $M_{c}=7\,\mathrm{M}_{\oplus}$. 
Since gas drag tends to smooth out the gap edges, they are
less steep in the semi-major axis distribution of smaller bodies, 
as illustrated in the left panels by less sharp transitions for smaller $R$.
Relative to the unperturbed (i.e., initial) distribution (dashed line in the
right panels), in the bottom panel of the figure, 
an excess of approximately $0.12$ and $0.25\,\mathrm{M}_{\oplus}$ worth of solids is 
collected at the inner and outer gap edge, respectively.
As mentioned above, the asymmetry is due to the fact that gas drag causes 
inward drift of small planetesimals. Mass tends to pile up at the outer edge 
of the gap, where drag is opposed by the core's perturbations, while these two
effects act in the same direction at the inner edge.
It is worth noticing that the calculations presented here do not include the effects
of planetesimals trapping into mean motion resonances with the planet.

\begin{figure*}
\centering%
 \resizebox{.80\hsize}{!}{\includegraphics[angle=00,clip]{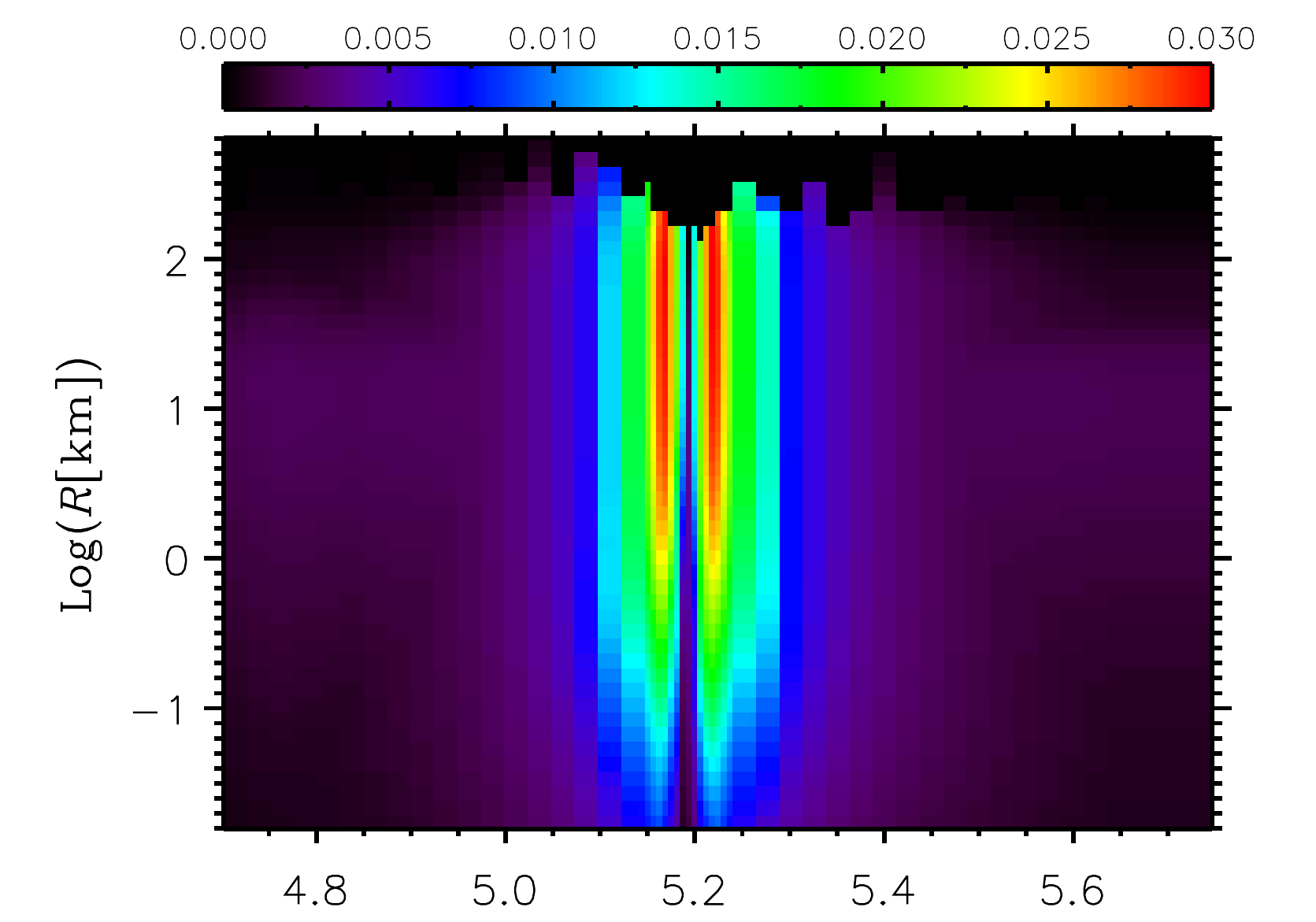}%
                                  \includegraphics[angle=00,clip]{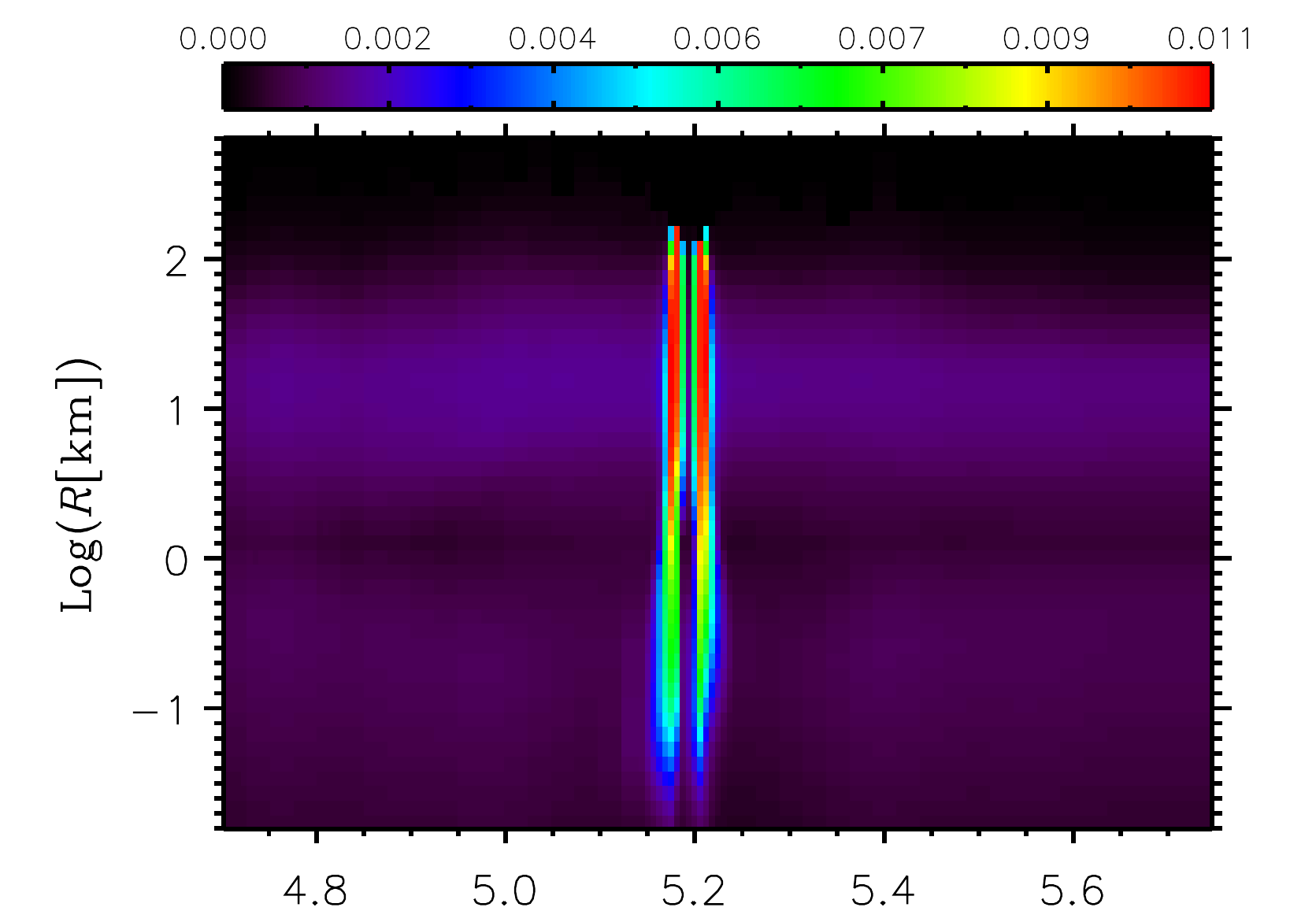}}
 \resizebox{.80\hsize}{!}{\includegraphics[angle=00,clip]{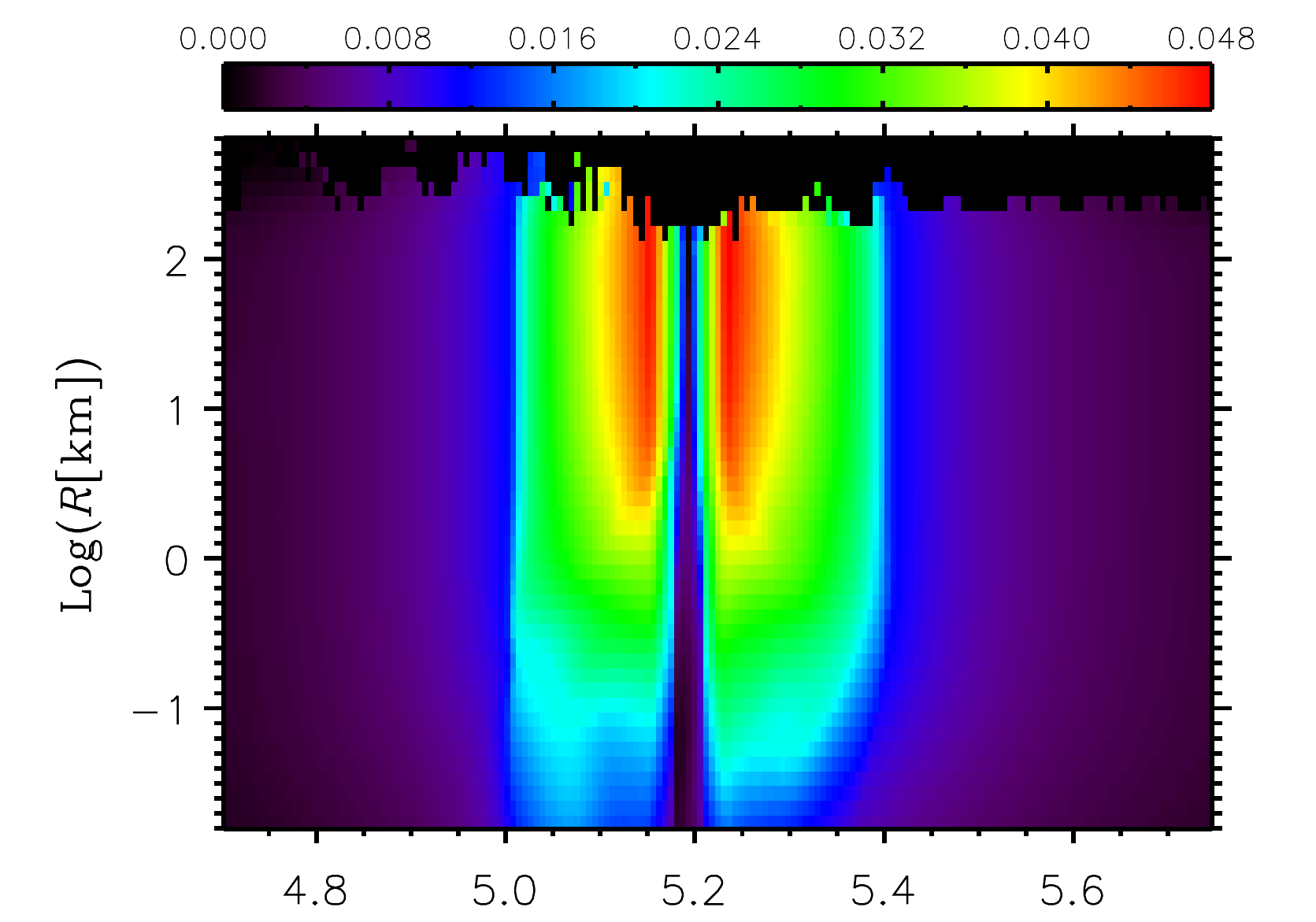}%
                                  \includegraphics[angle=00,clip]{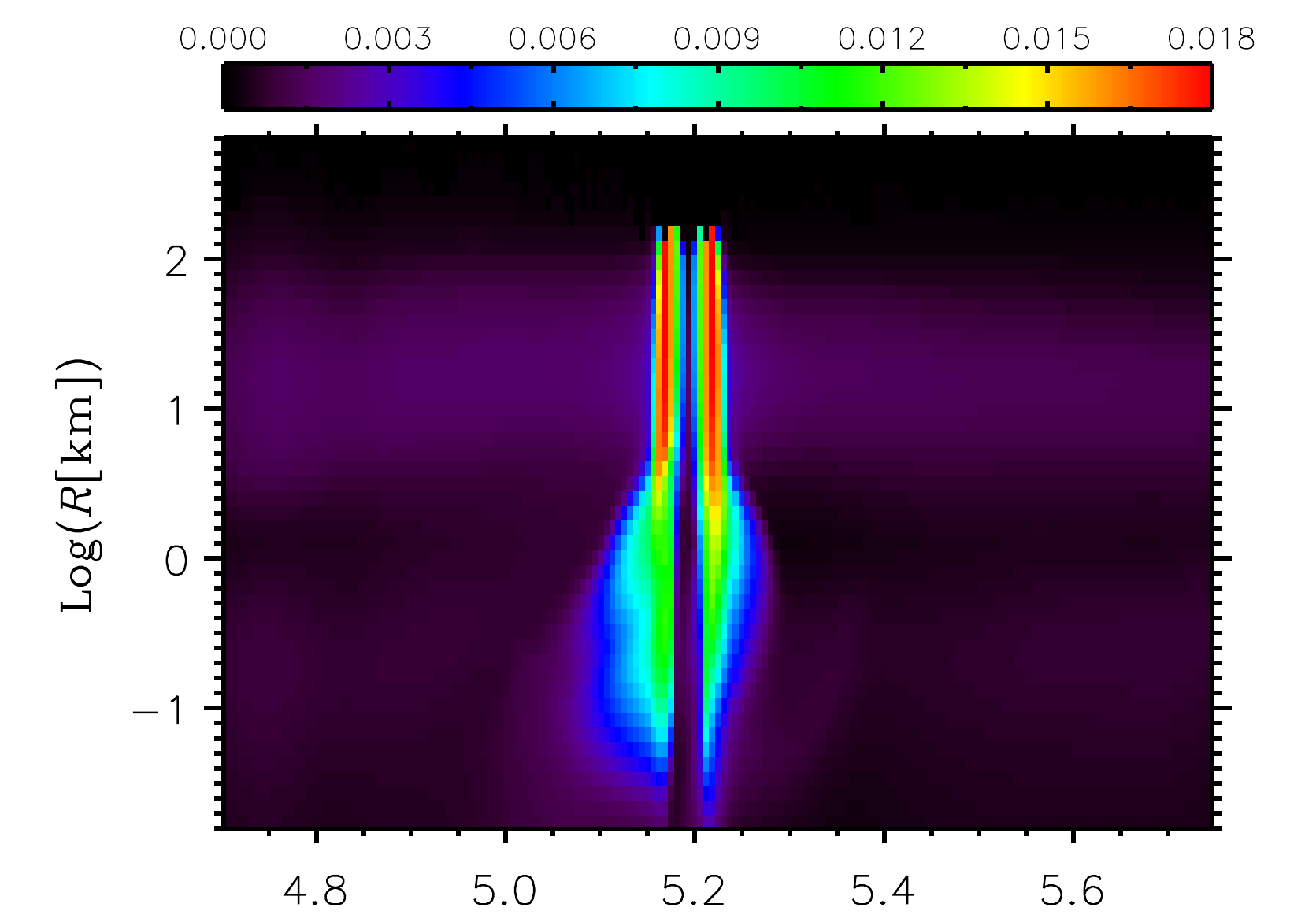}}
 \resizebox{.80\hsize}{!}{\includegraphics[angle=00,clip]{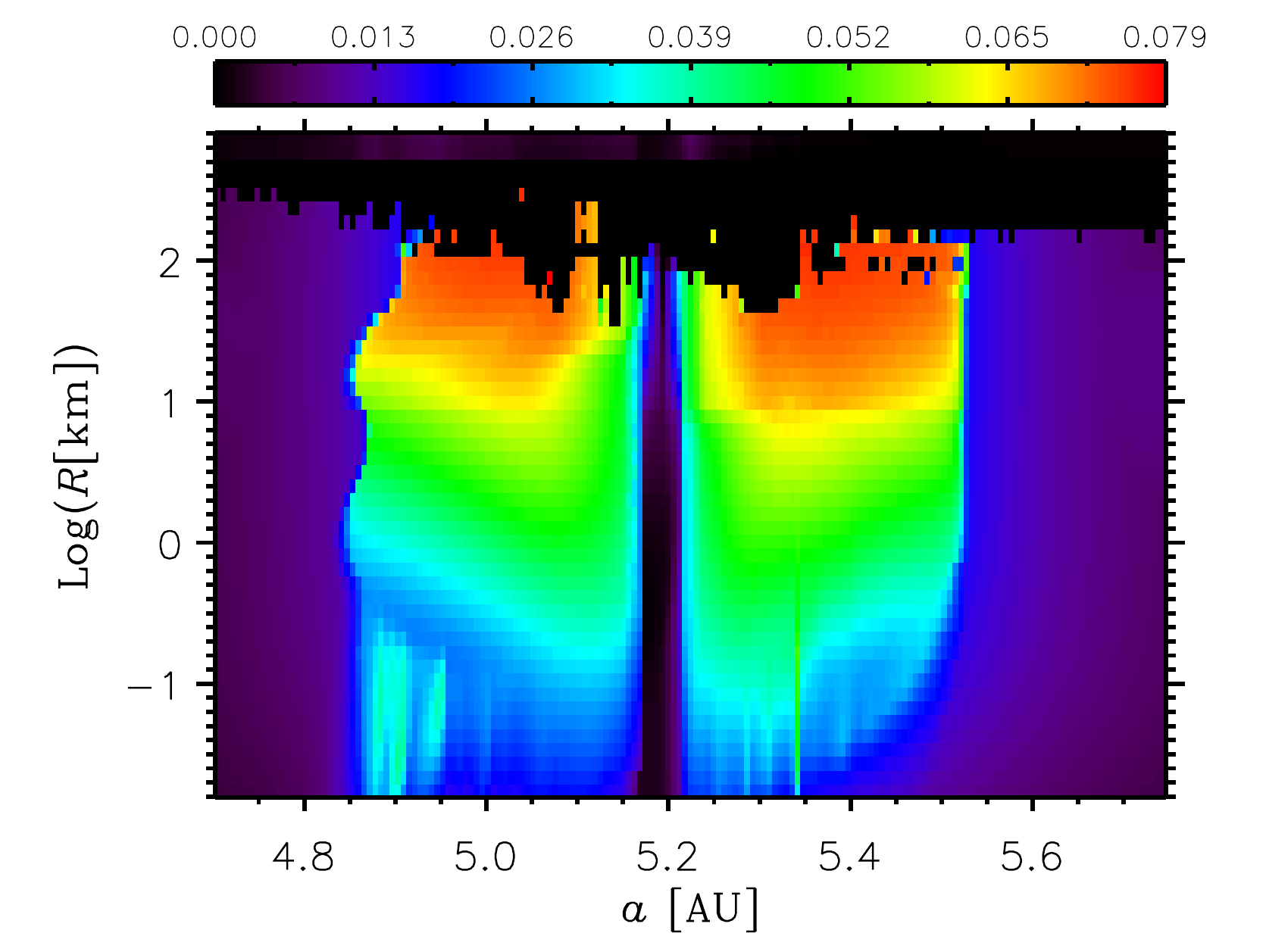}%
                                  \includegraphics[angle=00,clip]{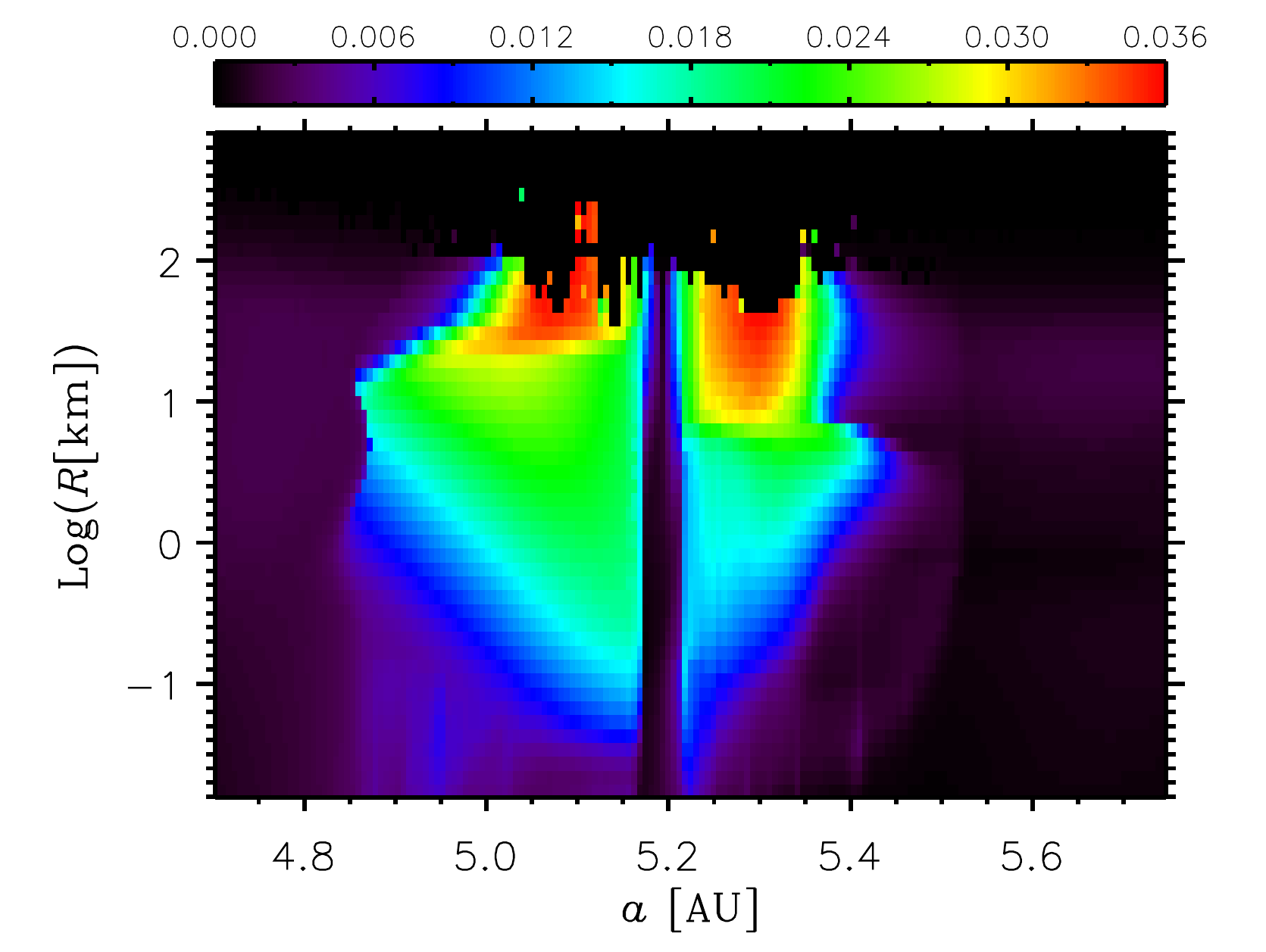}}
  \caption{%
               Distributions of mean orbital eccentricities (left) and of mean orbital
               inclinations (right) of planetesimals as a function of semi-major axis and radius.
               From top to bottom, the distributions are plotted when $M_{c}=1$, $3$, and 
               $7\,\mathrm{M}_{\oplus}$. 
               The right panels actually show the sine of the mean inclinations.
               }
  \label{fig:ei_maps}
\end{figure*}
The top and middle right panels of Figure~\ref{fig:maps} show double peaks
around the gap edges. The bodies in these regions have non-negligible 
eccentricities due to repeated synodical encounters with the core. 
This can be seen in the left panels of Figure~\ref{fig:ei_maps}, which show
the mean orbital eccentricities of planetesimals  versus semi-major axis and size.
The right panels of the figure show the mean orbital inclinations. We do not 
integrate individual orbits, but use a statistical estimate of the impact rate as 
a function of relative velocity and difference in semi-major axes from 
\citet{greenzweig1990} and \citet{greenzweig1992}. The double peak (or rather, 
a single valley in the region of enhanced surface density produced by shepherding) 
is due to our algorithm for collision probability, which is maximized for orbits 
that are tangential to that of the target body. Those with perihelia or aphelia 
at a distance equal to the core's semi-major axis are depleted more rapidly,
producing the local minima in surface density. It is not clear whether this is 
a real effect that would appear if the orbits were actually integrated, but 
it probably does not have a significant effect on the long-term evolution 
of the core mass.

\begin{figure}
\centering%
 \resizebox{\hsize}{!}{\includegraphics[angle=00]{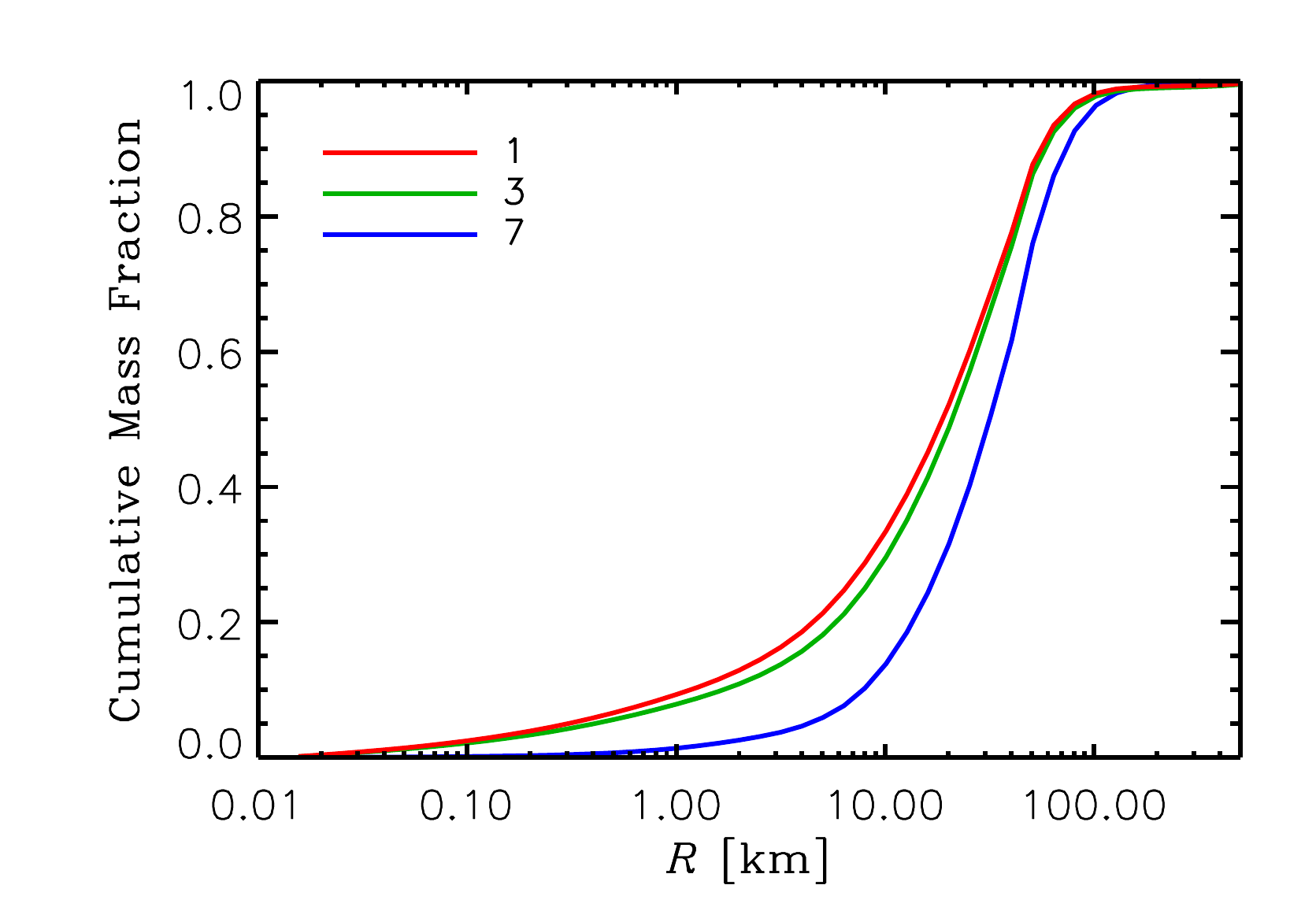}}
  \caption{%
               Cumulative mass of the planetesimal distributions
               obtained by integrating the data plotted in the left panels
               of Figure~\ref{fig:maps} over semi-major axis and then
               integrating from the minimum planetesimal radius to $R$. 
               The curves are 
               normalized to the total mass in the distribution. The three curves refer
               to the times when $M_{c}=1$, $3$, and $7\,\mathrm{M}_{\oplus}$,
               as indicated in the legend.
               }
  \label{fig:cum_mass}
\end{figure}
As anticipated above, most of the mass of the swarm is placed in bodies
with radii of tens of kilometers.
In the distributions of planetesimals shown in the left panels of 
Figure~\ref{fig:maps}, the single size bin that contains most mass, 
$10$\% of the total mass of the swarm at the time when 
$M_{c}=1\,\mathrm{M}_{\oplus}$ and $14$\% when 
$M_{c}=7\,\mathrm{M}_{\oplus}$, is that corresponding to radii
$45 \lesssim R\lesssim 55\,\mathrm{km}$. 
In the distributions illustrated in the upper two panels, as can be
seen from the cumulative mass fractions in Figure~\ref{fig:cum_mass},
less than $9$\% of the mass is in the form of planetesimals smaller than 
$1\,\mathrm{km}$ in radius, $<25$\% is in planetesimals with radii between 
$1$ and $10\,\mathrm{km}$, and $<2$\% is in planetesimals larger than
$100\,\mathrm{km}$ in radius.
Solids mass tends to be transferred to larger bodies as the swarm evolves. 
In the bottom panel of Figure~\ref{fig:maps}, only about $13$\% of the 
total mass is accounted for by planetesimals with radii $R \lesssim 10\,\mathrm{km}$ 
and about $7$\% is accounted for by planetesimals larger than $100\,\mathrm{km}$ 
in radius (see Figure~\ref{fig:cum_mass}).

\citet{pollack1996} assumed that the planet's feeding zone had a half-width
of $\sqrt{12\,R^{2}_{\mathrm{H}} + e^{2} a^{2}}$, and that the planetesimals were distributed
uniformly over the feeding zone. In our radially resolved simulations, the 
feeding zone is of comparable width, but the surface density is not uniform.
Planetesimals near the edges of the feeding zone (more than about 
$2.4\,R_{\mathrm{H}}$
from the core's orbit) require many synodic encounters with the core 
before their eccentricities become large enough to cross its orbit and
collide with it. Bodies whose semi-major axes are within 
$\sim 1\,R_{\mathrm{H}}$ of the core are in 
horseshoe or Trojan orbits, and are also initially unable to make close 
approaches to the core. Our multi-zone code can resolve the structure of the 
feeding zone. For a $5\,\mathrm{M}_{\oplus}$ core, the half-width is 
$\sim 0.3\,\mathrm{AU}$; we
typically use a width of $\sim 0.005\,\mathrm{AU}$ for each zone of semi-major axis. 
The initial 
rate of mass gain of the core is high due to shear-dominated runaway growth
as it sweeps up planetesimals in the range $\sim 1.8$-- $2.4\,R_{\mathrm{H}}$, 
but decreases as this favored region of the feeding zone becomes depleted. The 
phase of rapid growth lasts until the core reaches about one-third of the isolation
mass. Growth then continues at a slower rate, as planetesimals diffuse into 
the favored region of the feeding zone by collisions and nebular gas drag. In 
test simulations that do not include effects of a gaseous envelope captured 
by the core, this rate is so slow that the core cannot reach even half the isolation 
mass in Equation~(\ref{eq:Miso}) during the lifetime of the nebula. 

\subsection{Growth of a bare core}
\label{sec:GBC}
\begin{figure}
\centering%
 \resizebox{\hsize}{!}{\includegraphics[angle=00]{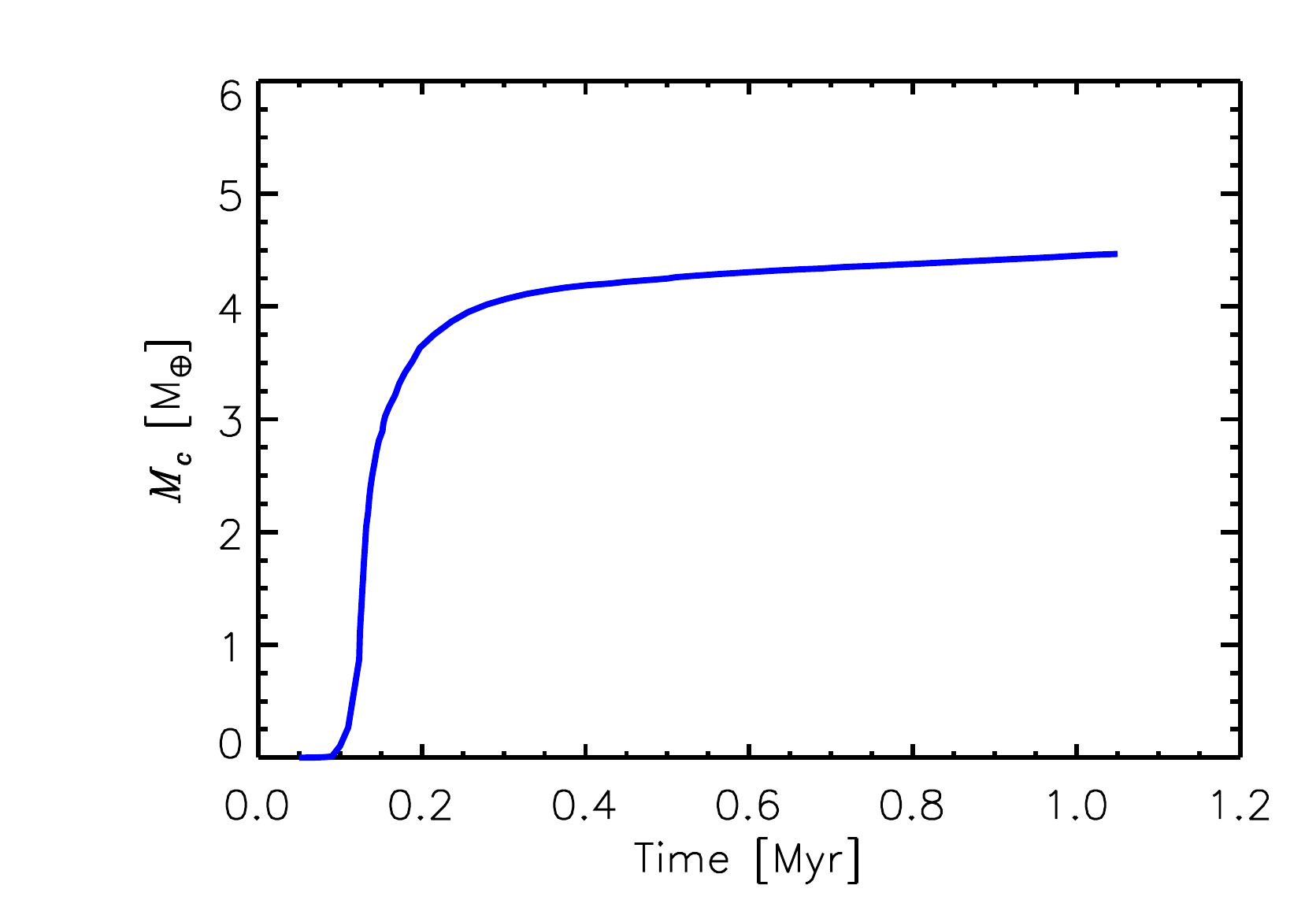}}
  \caption{%
               Core mass versus time in a simulation that
               does not take into account the enhancement of the 
               cross-section for collisional capture of planetesimals caused by 
               the gaseous envelope bound to the core. In $1\,\mathrm{Myr}$, 
               the core grows to a mass that is only about $40$\% of that expected 
               according to the standard concept of isolation mass (Equation~\ref{eq:Miso}).
               The maximum accretion rate is 
               $\approx 1.3\times10^{-4}\,\mathrm{M}_{\oplus}\,\mathrm{yr}^{-1}$.
               Toward the end of the calculation,  $\dot{M}_{c}$ is about 
               $3\times 10^{-7}\,\mathrm{M}_{\oplus}\,\mathrm{yr}^{-1}$ and
               a steeply declining function of $M_{c}$.
               }
  \label{fig:mc_bare}
\end{figure}
Figure~\ref{fig:mc_bare} shows the core mass versus time 
for our nominal simulation for the growth of a core \emph{without} the gaseous envelope.
Based on the initial mass-doubling timescale, it is assumed that the growth time 
of the initial seed body, from a size comparable to the largest bodies in
the initial swarm, is about $5\times 10^{4}$ years. 
Therefore, we set the time equal to $5\times 10^{4}$ years at the beginning 
of the simulation. 

In this model, the cross-section for accretion of planetesimals is
equal to the geometrical cross-section of the core.
The maximum rate of accretion, 
$\dot{M}_{c}\approx 1.3\times 10^{-4}\,\mathrm{M}_{\oplus}\,\mathrm{yr}^{-1}$, 
occurs when the core mass is around $1.5\,\mathrm{M}_{\oplus}$,
and the rate steadily declines afterward (except for sporadic fluctuations, 
see Figure~\ref{fig:naco}). 
The core only grows to a mass of about $4.4\,\mathrm{M}_{\oplus}$ in 
a little over $1\,\mathrm{Myr}$.
At the end of the calculation, the accretion rate is 
$\approx 3\times 10^{-7}\,\mathrm{M}_{\oplus}\,\mathrm{yr}^{-1}$.
Even if the accretion rate leveled off at this time, the core would reach 
a mass of $\approx 5\,\mathrm{M}_{\oplus}$ at a time of $2.7\,\mathrm{Myr}$ 
and would not attain half of the isolation mass in Equation~(\ref{eq:Miso})
until $4.3\,\mathrm{Myr}$ had elapsed.
In fact, at the end of the calculation, the decline of $\dot{M}_{c}$ as a
function of the core mass is so steep that, were it to continue at a comparable 
rate, a further $10$\% increment of $M_{c}$ would require an additional
$\sim 5\,\mathrm{Myr}$.

The effective mass limit of the planet in this case probably depends 
somewhat on the size distribution of the planetesimals, their impact strength, 
the fragmentation model, and the nebular gas density.
Additional details on the bare core calculation are presented in Section~\ref{sec:results}.

\subsection{Effects of a gaseous envelope on accretion of the core}
\label{sec:EGE}

The later stage of accretion can be aided by effects of a gaseous envelope
of hydrogen/helium captured from the nebula and retained by the planet's gravity.
The core begins to acquire an envelope significantly denser than the
surrounding nebular gas when it attains a mass typically 
$\gtrsim 0.1\,\mathrm{M}_{\oplus}$.
We compute the mass and structure of such an envelope by the 
method described in the next section. In our simulation with an envelope,
the cross-section of the core for planetesimal capture during close encounters
is augmented by drag exerted by the denser gas of the surrounding envelope. The 
cross-section is calculated self-consistently by solving for the structure of a 
quasi-hydrostatic envelope that grows in mass as it contracts. The capture 
cross-section for planetesimals becomes a function of both size and approach 
velocity. Small planetesimals that encounter the denser gas surrounding
the core are braked by aerodynamic drag, and may also be disrupted by
dynamical pressure, while larger bodies may pass through the outer region of
the envelope without being affected much. By the method described in the next
section, we find that initially the effective collision radius of the largest 
planetesimals, with diameters of a few hundred km, is comparable to the
physical radius of the solid core, $R_{c}$. 
But once the envelope mass grows beyond
$\sim 10^{-3}\,\mathrm{M}_{\oplus}$, the capture radius of
the largest bodies can become significantly larger than $R_{c}$.
For the smallest bodies, a few tens 
of meters in size, the effective collision radius may exceed $R_{c}$ by an order 
of magnitude or more. While the swarm is flat compared with the Hill radius, 
the collision radius for bodies entering the Hill sphere is much smaller,
so the collision cross-section varies as its square. Thus, the presence of
the envelope can increase the accretion rate of small bodies onto the 
core by two, or more, orders of magnitude.
  
The presence of the envelope also affects the gravitational stirring rate. 
Its mass added to that of the solid core increases the effective stirring.
In the early stages of core growth, the envelope mass is only a small
fraction of the total, and has little direct effect. However, there is another
subtle effect, as the stirring rate depends on the closest approach distance
that results in deflection without collision. The enhanced collision
cross-section increases the lower limit for such encounters, decreasing
the stirring rate, particularly for the smaller planetesimals. Another way
to describe this effect is that the small planetesimals that approach the
core are selectively removed by accretion, rather than stirred by such
encounters. This process effectively diminishes the stirring rate,
cooling the swarm. We compute the envelope structure and cross-section for 
planetesimal capture for each size bin, as outlined in the following sections.

\subsection{Envelope structure calculation}
\label{sec:ESC}

The envelope surrounding the growing core is described by the
typical equations of stellar structure \citep[e.g.,][]{kippenhahn2013},
with the additional energy source provided by the gravitational energy 
of incoming planetesimals.
These equations are solved using the 1-D code described in 
\citet{pollack1996,bodenheimer2000b,hubickyj2005}, with
modification described below in Section~\ref{sec:CRC}.
The gaseous envelope is supposed to have a protosolar ratio of
hydrogen and helium with a small admixture of heavier elements.
In the following, we shall refer to this part of the model simply as the 
envelope evolution calculation, as opposed to the core accretion calculation
described above.

Along with the module for the calculation of the envelope structure, 
the code includes three other main modules. The first module 
deals with the calculation of the interaction of the accreted planetesimals
with the gas in the envelope, including the computation of trajectories 
and ablation rates of the planetesimals as they travel through the envelope.
The second module calculates, in a self-consistent fashion, the opacity
at each depth in the envelope, accounting for sedimentation and 
coagulation of dust and small grains that are released in the envelope 
by ablating planetesimals, as described by \citetalias{naor2010}.
The  third module performs the calculation of the accretion of
nebular gas on the envelope, 
which includes the calculation of the limiting rates\footnote{These
limiting rates, at a heliocentric distance of $5.2\,\mathrm{AU}$, 
typically set in only when the planet mass grows beyond 
several dozen Earth masses.} 
of accretion at which the nebula can supply gas to the planet 
as described by \citet{lissauer2009}. 

The location of the inner boundary of the envelope is at the core radius, 
$R_{c}$. Prior to the phase of rapid envelope contraction (or runaway
gas accretion), the location of the outer boundary of the envelope, i.e., 
the planet radius $R_{p}$, is required to match the accretion radius. 
Based on thermal escape considerations and 
3-D hydrodynamics disk-planet interaction calculations, 
the inverse of the accretion radius, $1/R_{A}$, is set equal to the sum 
\begin{equation}
\frac{1}{R_{A}}=\frac{1}{R_{\mathrm{B}}}+\frac{4}{R_{\mathrm{H}}},
\label{eq:RAinv}
\end{equation}
where $R_{\mathrm{B}}$ and $R_{\mathrm{H}}$ are the Bondi and 
Hill radius, respectively. 
The factor $4$ in the equation comes from an estimate of the volume
in which the 3-D flow is bound to the planet \citep{lissauer2009}.
During this stage, density and temperature at $R_{p}$
are those of the ambient nebula gas, $\rho_{\mathrm{neb}}$ and
$T_{\mathrm{neb}}$, respectively.
 
\subsection{Planetesimal-envelope interaction and capture radii calculation}
\label{sec:CRC}

Planetesimals that travel through the hydrogen/helium envelope can lose a significant 
amount of kinetic energy if acted upon by a large enough drag force. 
The rate of change of the kinetic energy of a planetesimal as a result of 
the work performed by gas drag is
\begin{equation}
M v \dot{v}=-\frac{1}{2}\pi R^{2} \rho v^{3},
\label{eq:fric}
\end{equation}
where $R$ is the planetesimal radius, $v$ its velocity
relative to the gas, $M$ the planetesimal mass, and $\rho$
the gas density. The drag coefficient is assumed to be 1.
By solving for the velocity, we
find that $1/v=\pi R^{2} \rho t/(2 M)$ plus a constant. 
This implies that a $30$\% 
loss in velocity, or a $50$\% loss in the kinetic energy,
requires that the planetesimal travel through a mass of gas, 
$\pi R^{2} \rho v t$, about equal its own mass, $M$. 
Considering that the interaction occurs over a length of at most
$\sim R_{p}$, this condition also requires that the gas-to-planetesimal 
density ratio exceed the ratio $\sim R/R_{p}$.
These simple arguments neglect the increase of the gas density within 
the planet's envelope and the mass reduction of the planetesimal through 
ablation. Therefore, they underestimate the radius of captured bodies.
In fact, the condition $\pi R^{2} \rho v t \approx M$ is clearly aided by an
increasing $\rho$ and a  decreasing $M$.  

The numerical module that calculates the interaction between
an incoming planetesimal and the core's envelope was extended
to allow for the application to a size distribution of planetesimals.
The equations of motion account for the gravity force of the core
and the envelope, and for gas drag following the formalism of
\citet{podolak1988}. A sequence of trajectory integrations with 
impact parameter varying from zero to $R_{p}$ is attempted.
The integration also takes into account the mass loss of 
a planetesimal as it dissolves via ablation.
The effective collision radius, or capture radius $R_{\mathrm{capt}}$, 
is given by the largest 
impact parameter for which the planetesimal does not have enough 
(relative) kinetic energy to escape into heliocentric orbit, after one pass 
through the envelope.

After the determination of $R_{\mathrm{capt}}$, the calculation 
proceeds by performing a number of trajectory integrations, 
with impact parameter varying between zero and $R_{\mathrm{capt}}$, 
in order to determine and record the ablation history of a planetesimal,
or whether it breaks up before hitting the core, for each impact parameter.
This is done for all planetesimal sizes in the distribution, following 
the approach of \citet{podolak1988} to calculate the thermal balance
at the surface of the body and the ablation rates. 
Along its path in the envelope, a planetesimal can be completely 
ablated, break up, or hit the core. Averages over the various
trajectories (with differing impact parameter) provide the mean energy 
and mass deposited in the envelope at each depth.
The deposition of energy contributes to the energy budget of each 
envelope layer, whereas the
mass deposited represents the input for the dust sedimentation 
and coagulation
calculation, which includes the calculation of dust opacity. 
The ablated mass is supposed to eventually settle at the bottom of the 
envelope and is added to the core mass.
Notice, however, that this assumption is valid for the rocky component 
of the solids, while ices can dissolve in the envelope \citep{iaroslavitz2007}.
Future work will take into account the effects of the dissolved ice in the 
envelope outer layers.

The numerical integrator for the calculation of trajectories 
was upgraded to a fifth-order Dormand-Prince 
method with an adaptive step-size control based on the global accuracy 
of the solution \citep{hairer1993}. Convergence of the solution is
obtained within a relative tolerance of $10^{-5}$ or an absolute
tolerance of $10^{-10}$, whichever is achieved first.

\subsection{Solids accretion calculation with envelope-enhanced capture radii}
\label{sec:SAEEC}

The accretion of the core from the planetesimal swarm begins when 
$M_{c}=10^{-4}\,\mathrm{M}_{\oplus}$. Gas can become bound to the
core only when the escape velocity from the core exceeds the mean
thermal velocity of the ambient gas. Otherwise stated, it is necessary that
the Bondi radius $R_{\mathrm{B}}$ is sufficiently larger than the radius
of the core. At Jupiter's orbital distance 
($T_{\mathrm{neb}}\approx 120\,\mathrm{K}$), 
this condition typically requires a core whose mass is
$\sim 0.1\,\mathrm{M}_{\oplus}$.
However, around such a small core, most of the envelope would have 
a density on the order of the nebula density, and therefore 
$R_{\mathrm{capt}}\sim R_{c}$ for all but meter-size planetesimals.
Consequently, at the beginning of the evolution, we neglect the effects 
of the tenuous atmosphere that may form around the low-mass core
($M_{c}\le1\,\mathrm{M}_{\oplus}$).

We initialize the model as follows.
Let us indicate with $i=1,\ldots,N$ the size bins in the swarm of 
planetesimals.
We follow the growth of the seed body until 
it becomes a planetary core with mass $M_{c}=1.1\,\mathrm{M}_{\oplus}$
(see Figure~\ref{fig:mc_bare}),
applying capture radii $R_{\mathrm{capt}}(i)=R_{c}$ and generating
the solids' accretion rates, $\dot{M}_{c}(i)$ (see Section~\ref{sec:EPS}).
At this point, we start the calculation of the envelope structure
(the envelope mass is $M_{e}\sim 10^{-5}\,\mathrm{M}_{\oplus}$),
applying the accretion rates $\dot{M}_{c}(i)$ computed at
$M_{c}=1.1\,\mathrm{M}_{\oplus}$ and producing the envelope-enhanced
capture radii, $R_{\mathrm{capt}}(i)$, for each planetesimal radius, 
according to the procedure outlined in Section~\ref{sec:CRC}.
The capture radii are also ``centered'' in core mass  at
$M_{c}=1.1\,\mathrm{M}_{\oplus}$.
The envelope-enhanced capture radii are then used to update 
and evolve the core accretion 
calculation from $M_{c}=1.0$ to $1.2\,\mathrm{M}_{\oplus}$.
As mentioned above, the envelope mass is also taken into account
in the core accretion calculation.
At the end of the initialization procedure, we have solids' accretion rates
computed for $M_{c}=1.2\,\mathrm{M}_{\oplus}$, based on capture radii 
computed for $M_{c}=1.1\,\mathrm{M}_{\oplus}$. 
The envelope evolution and core accretion calculations are staggered
in core mass by an amount equal to $\Delta M_{c}=0.1\,\mathrm{M}_{\oplus}$.

The overall calculation proceeds in a step-wise fashion, by advancing the
envelope evolution and core accretion calculations so that the core mass
increases by $\Delta M_{c}=0.2\,\mathrm{M}_{\oplus}$ in each step,
and exchanging data ($\dot{M}_{c}(i)$, $R_{\mathrm{capt}}(i)$, and
$M_{e}$) between the two calculations at the end of each step. 
As a result, in advancing from $M_{c}$ to $M_{c}+\Delta M_{c}$,
each calculation employs information from the other calculation computed 
at $M_{c}+\Delta M_{c}/2$.

We find that this procedure yields maximum variations of $\dot{M}_{c}(i)$,
from one step to the next, of less than about $25$\% and even smaller 
maximum variations of $R_{\mathrm{capt}}(i)$. As the planet mass grows, 
the gas accretion rate, $\dot{M}_{e}$, increases, eventually causing 
significant variations of the capture radii from one step to the next
(smaller bodies may be especially sensitive to variations of the envelope
mass, as explained in Section~\ref{sec:EGE}). 
Therefore, within an envelope evolution step, if the maximum variation 
of $R_{\mathrm{capt}}(i)$ with respect to $i$ exceeds $15$\%, we update 
$\dot{M}_{c}(i)$ by repeating the last core accretion step and applying the 
latest values computed for $R_{\mathrm{capt}}(i)$ and $M_{e}$.
In addition, once gas accretion dictates planet growth, that is 
$\dot{M}_{e}>\dot{M}_{c}=\sum^{N}_{1}\dot{M}_{c}(i)$, the data
exchange between calculations is executed at mass intervals 
$\Delta M_{c}=0.05\,\mathrm{M}_{\oplus}$.
However, the condition $\dot{M}_{e}>\dot{M}_{c}$ also marks 
the end of Phase~1 and the beginning of Phase~2 of the planet evolution
(see Section~\ref{sec:introduction}), which will be described in 
a forthcoming paper.

\subsection{Model improvement via a predictor-corrector scheme}
\label{sec:CM}

The determination of the accretion rates of solids is affected by 
stochastic variations, for example, of the number of impacts on the core, 
which may add a random component to the physical variations of 
$\dot{M}_{c}(i)$, from one step to the next, in the scheme outlined above.
In order to compensate for such effects, we can consider the calculation 
described so far as a ``predictor'' approximation of a predictor-corrector 
scheme.
 
The corrected calculation proceeds as follows.
Indicating with $\dot{M}^{k}_{c}(i)$, for $k=1, \ldots, K$, the values of 
$\dot{M}_{c}(i)$ for increasing total planet mass, $M^{k}_{p}$
($M_{p} =M_{c}+M_{e}$), 
we reduce stochastic variations over small changes of $M_{p}$ 
by introducing smoothed values of the accretion rate for each size bin
\begin{equation}
\langle\dot{M}_{c}\rangle^{k}_{i}=\frac{1}{3}%
\left[\dot{M}^{k-1}_{c}(i)+\dot{M}^{k}_{c}(i)+\dot{M}^{k+1}_{c}(i)\right].
\label{eq:dMcdt_smoo}
\end{equation}
The edge values are defined as
$\langle\dot{M}_{c}\rangle^{1}_{i}=[2\dot{M}^{1}_{c}(i)+\dot{M}^{2}_{c}(i)]/3$ and
$\langle\dot{M}_{c}\rangle^{K}_{i}=[\dot{M}^{K-1}_{c}(i)+2\dot{M}^{K}_{c}(i)]/3$.
Notice that the smoothing operator is globally conservative, in the sense that
\begin{equation}
\sum_{k=1}^{K}\langle\dot{M}_{c}\rangle^{k}_{i}=%
\sum_{k=1}^{K}\dot{M}^{k}_{c}(i),
\label{eq:dMcdt_con}
\end{equation}
as can be easily checked by direct substitution.

Equation~(\ref{eq:dMcdt_smoo}) is applied to all the accretion rates'
datasets used in the ``predictor'' calculation. We then build continuous 
functions of the planet mass 
$\langle\dot{M}_{c}\rangle_{i}=\langle\dot{M}_{c}\rangle_{i}(M_{p})$ 
by performing linear interpolations of the smoothed accretion rates, 
$\langle\dot{M_{c}\rangle}^{k}_{i}$.
The corrected calculation is obtained by restarting the envelope evolution
from the beginning ($M_{c}=1.1\,\mathrm{M}_{\oplus}$) and using
the functions $\langle\dot{M}_{c}\rangle_{i}$ as accretion rates.

\begin{figure}
\centering%
 \resizebox{\hsize}{!}{\includegraphics[angle=00]{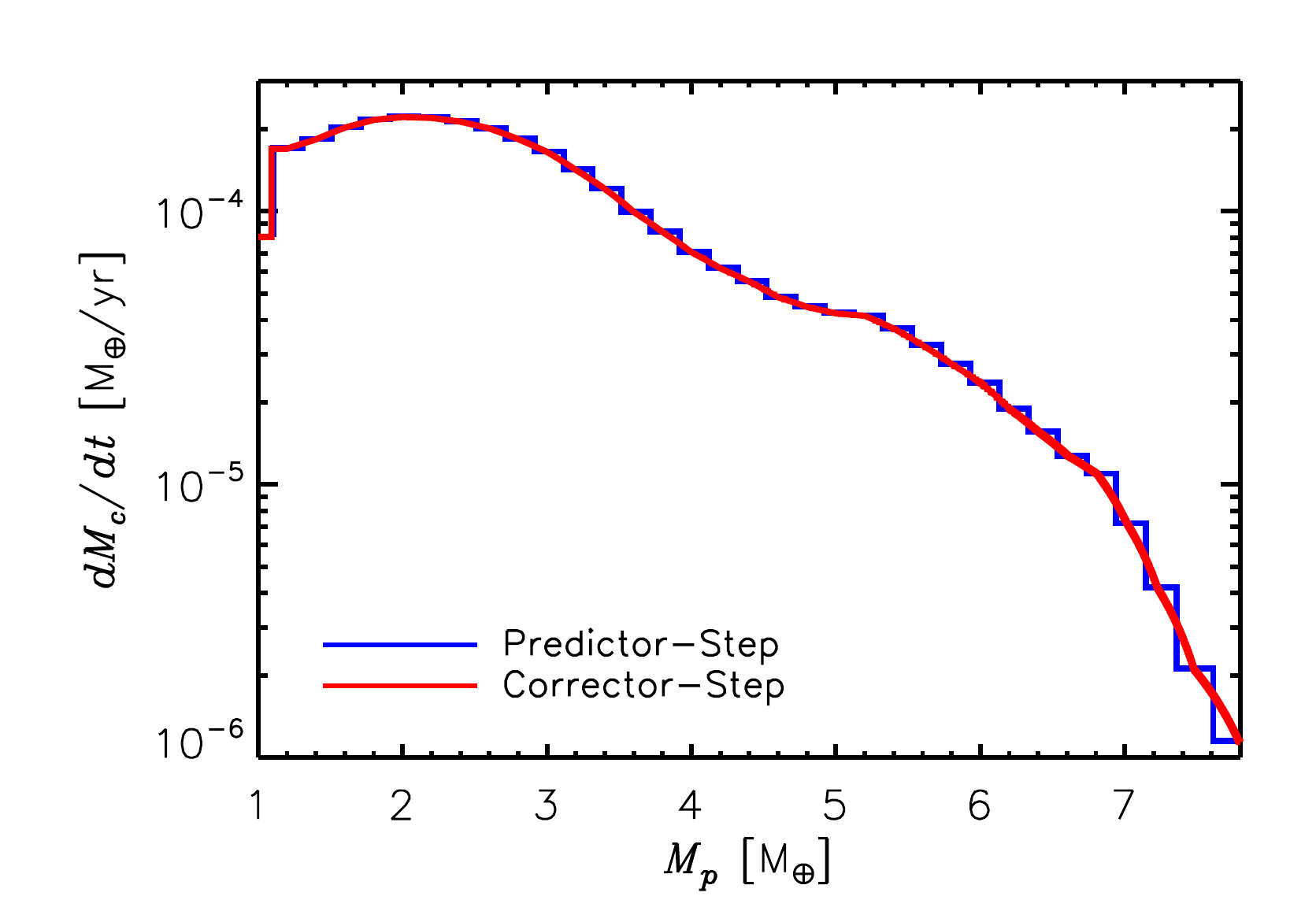}}
  \caption{%
               The plot shows the accretion rate of solids $\dot{M}_{c}=\sum_{i}\dot{M}_{c}(i)$
               (step function) applied in the ``predictor'' calculation (see Section~\ref{sec:CM})
               versus the planet mass, $M_{p}=M_{c}+M_{e}$. The quantity $\dot{M}_{c}(i)$
               indicates the accretion rate for a given planetesimal radius in the size distribution.
               The smooth line represents
               the accretion rate  $\langle\dot{M}_{c}\rangle=\sum_{i}\langle\dot{M}_{c}\rangle_{i}$, 
               applied in the ``corrector'' calculation.
               }
  \label{fig:precor}
\end{figure}
Figure~\ref{fig:precor} shows the total accretion rate used in the ``predictor'' step, 
$\dot{M}_{c}$ (step function), as well as that used in the corrected calculation, 
$\langle\dot{M}_{c}\rangle$ (continuous function).
We find that the smoother behavior of the accretion rate in solids
leads to a smoother increase of the envelope mass, especially when
$\dot{M}_{c}$ changes by more than $\sim 30$\% from one step to the next.
We also find that the capture radii computed in the corrected calculation are
very similar to those used in the ``predictor'' calculation
(differences are within a few percent), for which reason a corrected version of 
the core accretion calculation is not performed.

In the rest of the paper, we will only refer to the corrected calculation.
Therefore, for ease of reading, the notation $\langle\dot{M}_{c}\rangle$ 
will be dropped, and the accretion rate of solids is simply denoted
as $\dot{M}_{c}$.

\section{Comparison of the method with previous work}
\label{sec:compa}

The analytic and statistical calculations of \citet{kobayashi2010,kobayashi2011} 
provide planetesimal accretion rates that include collisional fragmentation 
and a range of planetesimal sizes, with the details somewhat different 
from those used here. The atmospheric enhancement of the planetesimal 
capture cross-section is included, with a more approximate atmospheric
structure and effective capture radius than those described in 
Sections~\ref{sec:ESC} and \ref{sec:CRC}. The opacity
calculation does not include grain settling and coagulation but 
rather assumes interstellar grain opacities multiplied by 
an arbitrary factor. They found that a disk with ten times 
the solid surface density of the minimum-mass solar
nebula and an initial planetesimal size of $100\,\mathrm{km}$ 
(or larger) are required to generate a solid core of 
$10\,\mathrm{M}_{\oplus}$ in less than $10^{7}$ years 
between $5$ and $10\,\mathrm{AU}$. An additional
requirement is that the grain opacity be reduced by a factor 
$\approx 100$ compared with interstellar values.  

Independent simulations by \citet{benvenuto2009} and
\citet{fortier2009}, based on the method described in
\citet{fortier2007}, employ a detailed model envelope and
a capture radius calculation similar to ours. However the
core accretion rate is simplified in that it does not use
a statistical simulation with the calculation of the
evolution of the planetesimal size distribution, but rather
assumes a size distribution fixed in time or, alternatively,
a single size. The planetesimal velocities are
calculated according to an assumed balance between the
stirring by the embryo and the damping by gas drag. The
grain opacities are interstellar. The initial mass of the
embryo is about $10^{-2}\,\mathrm{M}_{\oplus}$. 
\citet{benvenuto2009} found that Jupiter, at a fixed distance 
of $5.5\,\mathrm{AU}$ in a disk with a solid surface density 
$\sigma_{Z} =11\,\mathrm{g\,cm^{-2}}$, about $3.3$ times 
that expected in the minimum-mass solar nebula, can form in less
than $1\,\mathrm{Myr}$ if most of the planetesimals are in the size
range $30$ to $100$ meters. However, if most of the planetesimal
mass is in the kilometer size range, formation times are longer,
about $6\,\mathrm{Myr}$. \citet{fortier2009} performed a similar 
calculation. 
For Jupiter at $5.2\,\mathrm{AU}$, with a single planetesimal size 
of $10\,\mathrm{km}$, the formation time is $2.4\,\mathrm{Myr}$ 
for a $\sigma_{Z}$ six times that of the minimum-mass solar nebula. 
If the size is increased to $100\,\mathrm{km}$, the formation time is 
longer: $3.2\,\mathrm{Myr}$ for a $\sigma_{Z}$ ten times that of 
the minimum-mass solar nebula.
Clearly, the details of the assumed planetesimal size
distribution have an important influence on the results
\citep{guilera2011}.

A completely different approach is taken by 
\citet{levison2010}. They do an N-body simulation starting
with $5$ embryos of $1\,\mathrm{M}_{\oplus}$ each,
situated between $4.5$ and $6.5\,\mathrm{AU}$, 
embedded in a disk of planetesimals with
solid surface density $6$ times that of the minimum-mass 
solar nebula. The simulations include embryo-disk tidal 
interactions, collisional damping, planetesimal-induced 
migration, embryo atmospheres, gas drag, and fragmentation.
In their models, it is difficult to grow embryos to $10$ Earth masses. 
The embryos open up a gap in the planetesimal disk in
their vicinity and do not grow to more than about
$2\,\mathrm{M}_{\oplus}$. In a few exceptional cases, the outer
embryos undergo rapid, planetesimal driven, outward
migration, and can accrete to up to $30\,\mathrm{M}_{\oplus}$.
Given the differences in various physical assumptions, the different
formation timescales between the calculations of \citeauthor{levison2010} 
and the calculation presented herein are not surprising. Perhaps, the most 
important differences concern the inclusion of embryo migration and the number 
of growing embryos. We do not allow the growing core to drift in response to 
the nebula torques and to core-planetesimals interaction whereas the inclusion 
of this effect is one of the main motivations driving \citeauthor{levison2010}'s 
investigation. In addition, while we introduce in the swarm an initial seed of 
$10^{-4}\,\mathrm{M}_{\oplus}$ (which is large enough to become the dominant 
embryo), they use $5$ closely spaced embryos that are much larger in mass. 
Moreover, the swarm dynamics and the gaseous envelope of the embryos are 
also modeled quite differently.

Over the past few years, the importance of the accretion of small,
centimeter-to-meter size, solids has been investigated
\citep[e.g.,][]{ormel2010,lambrechts2012,morbidelli2012,chambers2014}.
Because of their size, these bodies are subjected to a strong aerodynamic 
drag force. According to Equation~(\ref{eq:tau_drag}), corrected for an appropriate 
drag coefficient, the orbital radius of meter-size bodies at $\sim 5\,\mathrm{AU}$
would significantly shrink over the orbital period.
(Notice, however, that Equation~(\ref{eq:tau_drag}) is derived under the assumption
of a Keplerian-like orbit.)
Therefore, these fragments should be continuously replenished via collisional 
comminution of planetesimals in order to represent a persistent source of accretion.
Although such comminution is allowed in our simulation, the larger planetesimals 
have significant gravitational binding energy, and are not prone to disruption. 
As mentioned in Section~\ref{sec:PAC}, the total mass lost at small sizes 
($< 15\,\mathrm{m}$) is about $10$\% of the total mass of the swarm.
Even if the core accreted the fragments exterior of its orbit with a $100$\% efficiency,
the core mass would change only marginally (by about $10$\%).
For accretion of small solids to be important in our simulation, we should have postulated 
that the initial planetesimals were much more fragile and prone to disruption, and/or that 
small bodies were abundant in the outer nebula and supplied more mass to the feeding 
zone than was originally present. 
But these scenarios are beyond the scope of the present investigation.

Furthermore, \citet{chambers2014} performed simulations of the oligarchic
growth of giant planet cores including a populations of millimeter-to-meter size
particles, which are produced by the collisional cascade initiated by larger
($1$--$100\,\mathrm{km}$ size) planetesimals. 
He concluded that the importance for core
accretion of these small particles depends on a balance between the production
rate (via collisional comminution) and the loss rate (via drag-induced radial drift). 
If the parent population
of planetesimals is constituted of objects of $\sim 1\,\mathrm{km}$ in diameter, 
then the accretion rate in small particles is significant. However, if the parent
planetesimals are large, $\sim 100\,\mathrm{km}$ in diameter, then small
particles do not provide a major source of solids accretion.
In our calculation, most mass resides in large planetesimals, 
$\gtrsim 100\,\mathrm{km}$ in diameter (see Section~\ref{sec:EPS}). 
Therefore, according to these findings, 
millimeter-to-meter size particles should not contribute significantly to the
growth of the core in our calculation.

\section{Results}
\label{sec:results}

\begin{figure}
\centering%
 \resizebox{\hsize}{!}{\includegraphics[angle=00]{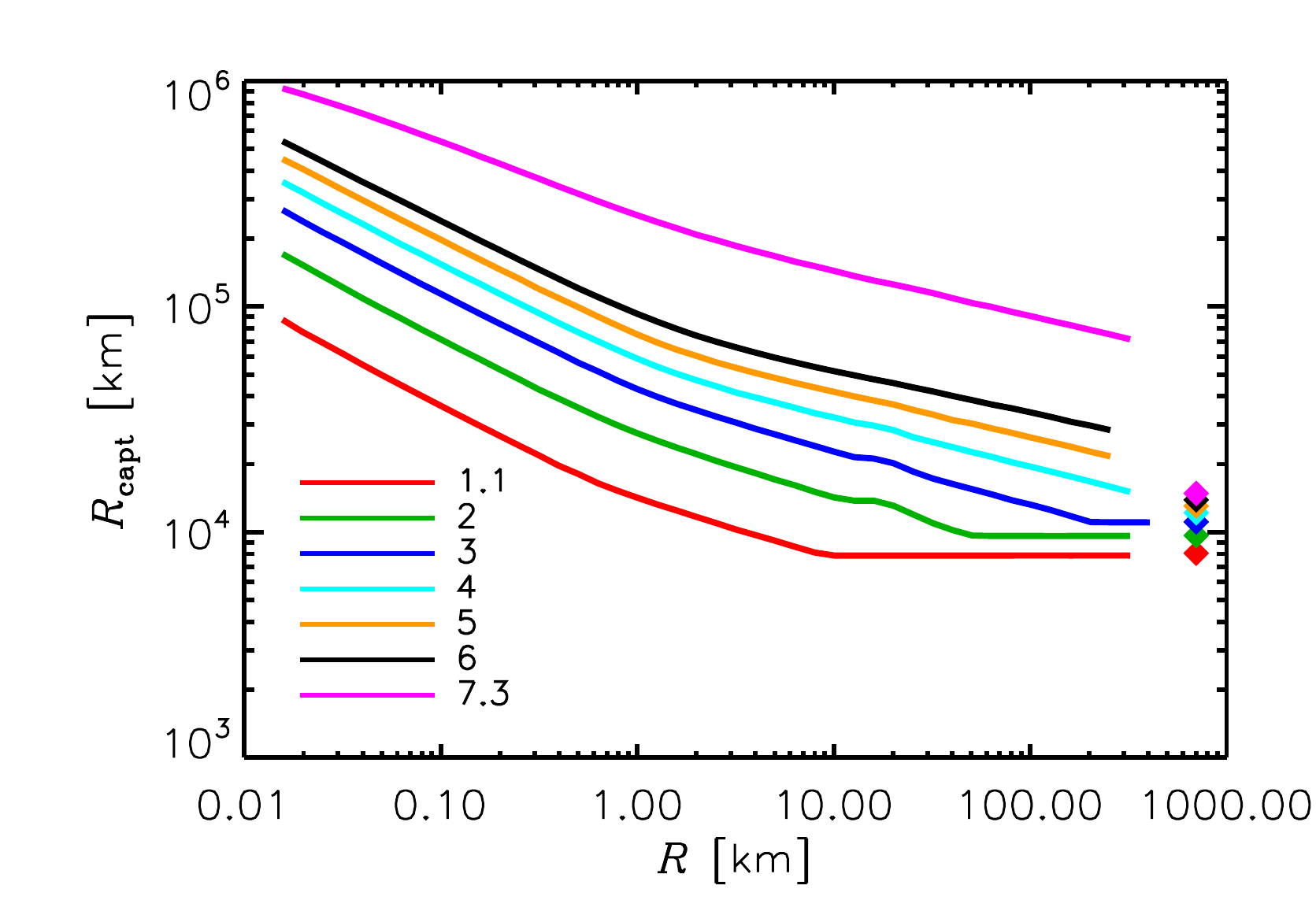}}
  \caption{%
               The envelope-enhanced capture radius for planetesimal accretion 
               versus planetesimal radius for different values of the core mass, 
               $M_{c}$, in units of $\mathrm{M}_{\oplus}$, as indicated in the legend.
               The symbols on the far right side of the diagram indicate the core radius, $R_{c}$.
               In the envelope around $1$, $2$, and $3\,\mathrm{M}_{\oplus}$ cores, larger
               planetesimals can reach the core and $R_{\mathrm{capt}}$ levels off to $R_{c}$. 
               The inflection in the curves, around $R\sim 10\,\mathrm{km}$
                for $M_{c}=2$ and $3\,\mathrm{M}_{\oplus}$, is caused by a regime 
                transition between complete ablation and disruption.
               }
  \label{fig:rcapt}
\end{figure}
As anticipated in the previous sections, the capture radius of the planet
for planetesimal accretion, $R_{\mathrm{capt}}$, is substantially enhanced 
by the presence of the gaseous envelope. Figure~\ref{fig:rcapt} shows
$R_{\mathrm{capt}}$ versus planetesimal radius, $R$, for different core masses.
At $\approx 1\,\mathrm{M}_{\oplus}$, the envelope mass is small, 
$\sim 10^{-5}\,\mathrm{M}_{\oplus}$, yet planetesimals up to a few
kilometers in radius are already affected by gas drag in the envelope
so that $R_{\mathrm{capt}}>R_{c}$.
In fact, when $R=4\,\mathrm{km}$, $R_{\mathrm{capt}}$ already exceeds 
the core radius by about $20$\%.
For $R>10\,\mathrm{km}$, $R_{\mathrm{capt}}$ is basically equal to $R_{c}$, 
as indicated by the symbols on the right side of the figure.
When $M_{e}$ grows to $10^{-4}\,\mathrm{M}_{\oplus}$
($M_{c}\sim 2\,\mathrm{M}_{\oplus}$), bodies up to 
$50\,\mathrm{km}$ in radius are affected by the envelope.
Once the envelope mass exceeds $\sim 10^{-3}\,\mathrm{M}_{\oplus}$ 
($M_{c}\sim 4\,\mathrm{M}_{\oplus}$),
the planet has a capture radius larger than the core radius for
essentially all planetesimals
($R_{\mathrm{capt}}/R_{c}\approx 1.31$ for $250$-km radius planetesimals 
and $\approx 1.24$ for $320$-km radius planetesimals).
Around a core mass of about $6\,\mathrm{M}_{\oplus}$ and  
$M_{e}\sim 4\times 10^{-3}\,\mathrm{M}_{\oplus}$, the cross-section of the planet
for accretion of $250$-km radius planetesimals is $4$ times as large
as the geometrical cross-section. 
When $M_{c}= 7.3\,\mathrm{M}_{\oplus}$ ($M_{e}\approx 0.15\,\mathrm{M}_{\oplus}$),
the cross-section for accretion of $320$-km radius planetesimals 
exceeds the geometrical cross-section by a factor of over $20$!

\begin{figure*}
\centering%
 \resizebox{\hsize}{!}{\includegraphics[angle=00]{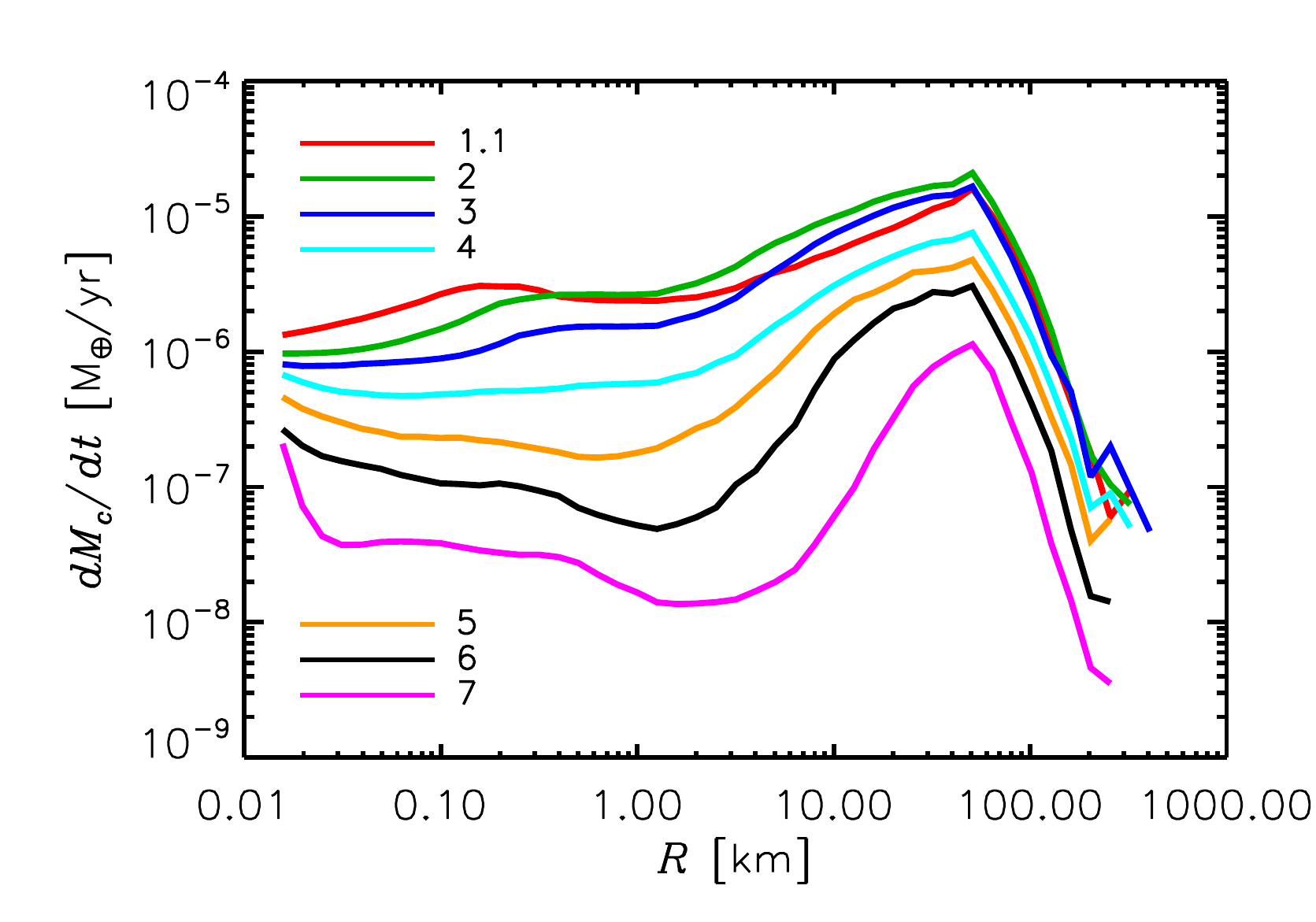}%
                                  \includegraphics[angle=00]{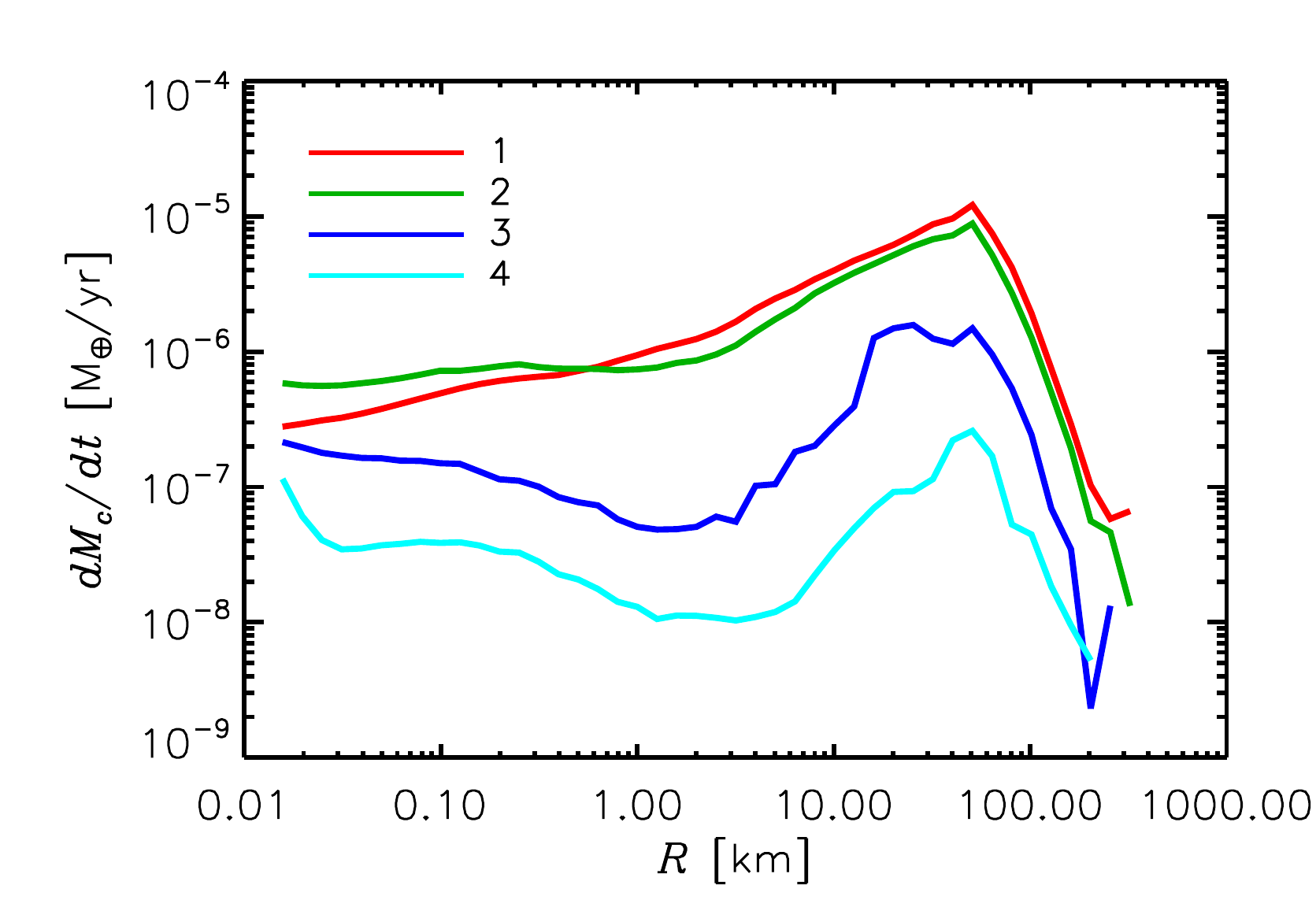}}
 \resizebox{\hsize}{!}{\includegraphics[angle=00]{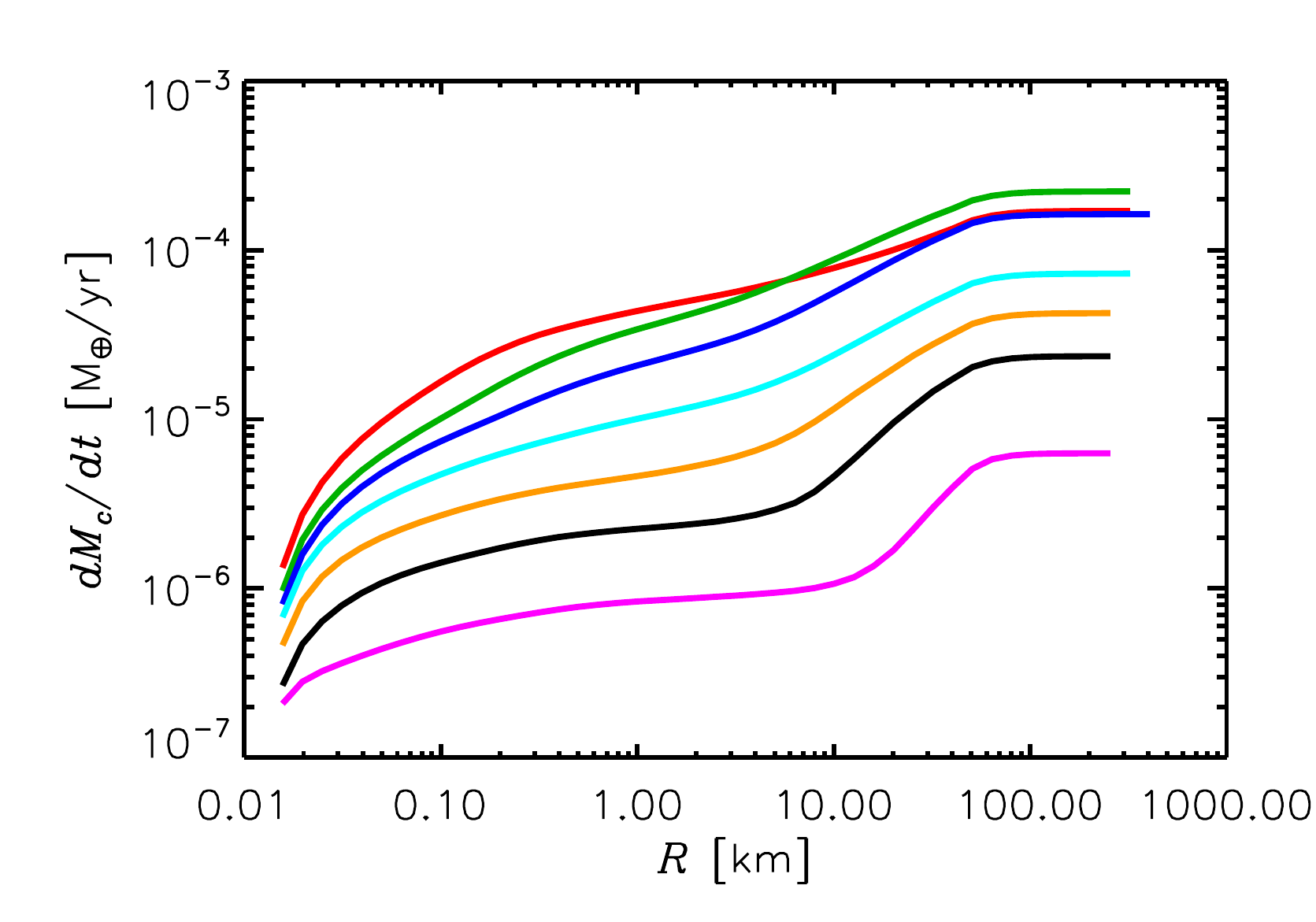}%
                                  \includegraphics[angle=00]{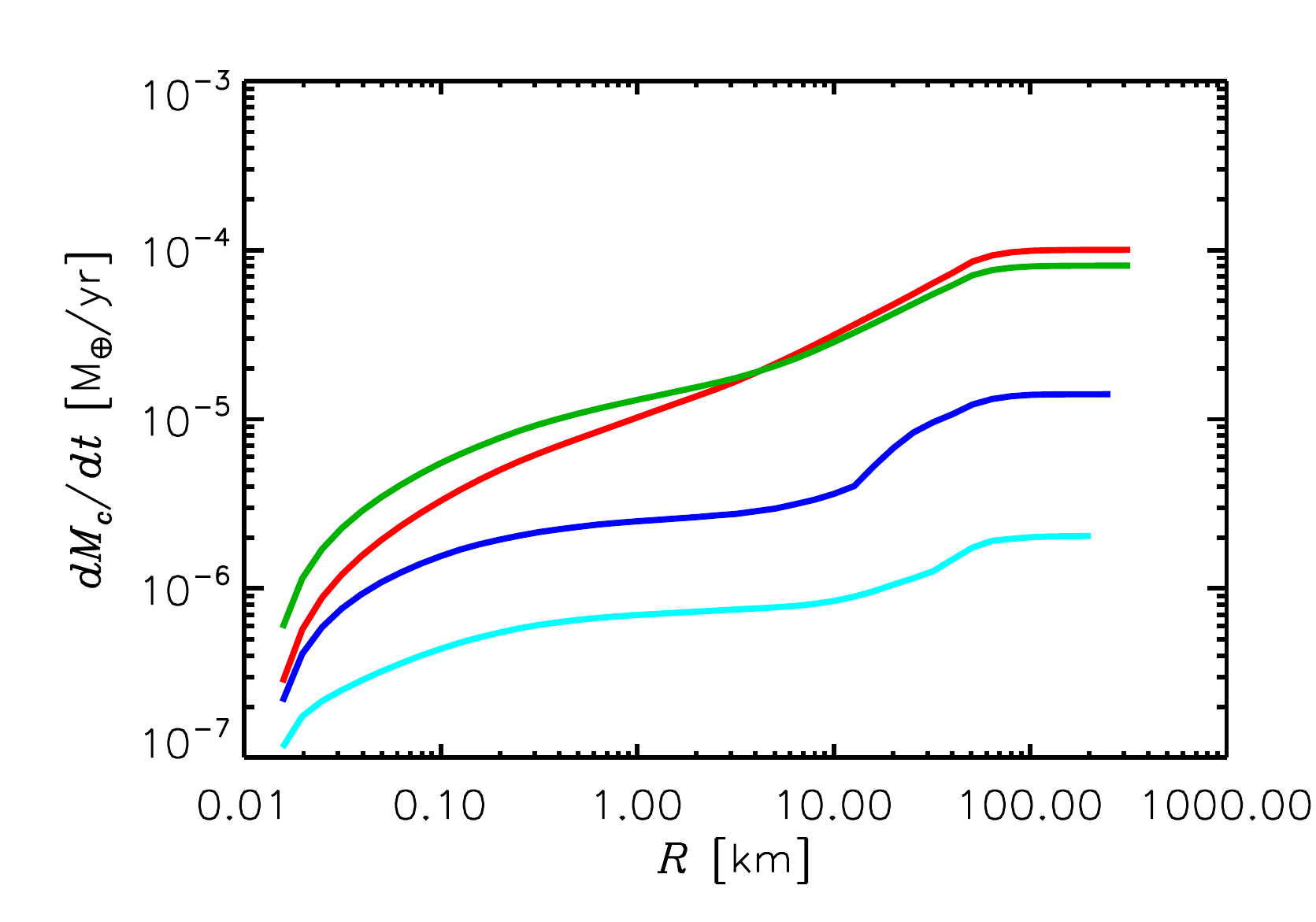}}
  \caption{%
               The top-left panel shows the accretion rate of solids as a function of 
               the planetesimal radius at various core masses, in units of 
               $\mathrm{M}_{\oplus}$, as indicated.
               The bottom-left panel shows the cumulative distribution of the
               accretion rate, at the same core masses, obtained from the curves
               in the top panel. For comparison, the same quantities are plotted 
               in the right panels ($M_{c}=1$, $2$, $3$, and $4\,\mathrm{M}_{\oplus}$) 
               for the calculation of the growth of a core that does not have 
               an envelope (see Section~\ref{sec:EPS}). Notice the large difference
               in $\dot{M}_{c}$ between the left and right panels, especially for
               $M_{c}>2\,\mathrm{M}_{\oplus}$, caused by gas drag in the growing 
               planet's envelope.
               }
  \label{fig:dmcdt}
\end{figure*}
The accretion rate of solids by the planet versus planetesimal radius is plotted
in Figure~\ref{fig:dmcdt} (top panels), along with the cumulative distributions
(bottom panels). Left and right panels correspond, respectively, to the calculation 
with and without envelope. 
In the latter calculation, $R_{\mathrm{capt}}=R_{c}$ for all planetesimal in the swarm.
Around the lowest core masses ($M_{c}\approx 1\,\mathrm{M}_{\oplus}$), 
$\dot{M}_{c}=\dot{M}_{c}(R)$ is affected by the planet's envelope mostly at small 
planetesimal radii, which results in accretion rates larger by factors of $5$ at 
$R\approx 15\,\mathrm{m}$ and about $2.5$ at $R\approx 1\,\mathrm{km}$.

\begin{figure}
\centering%
 \resizebox{\hsize}{!}{\includegraphics[angle=00]{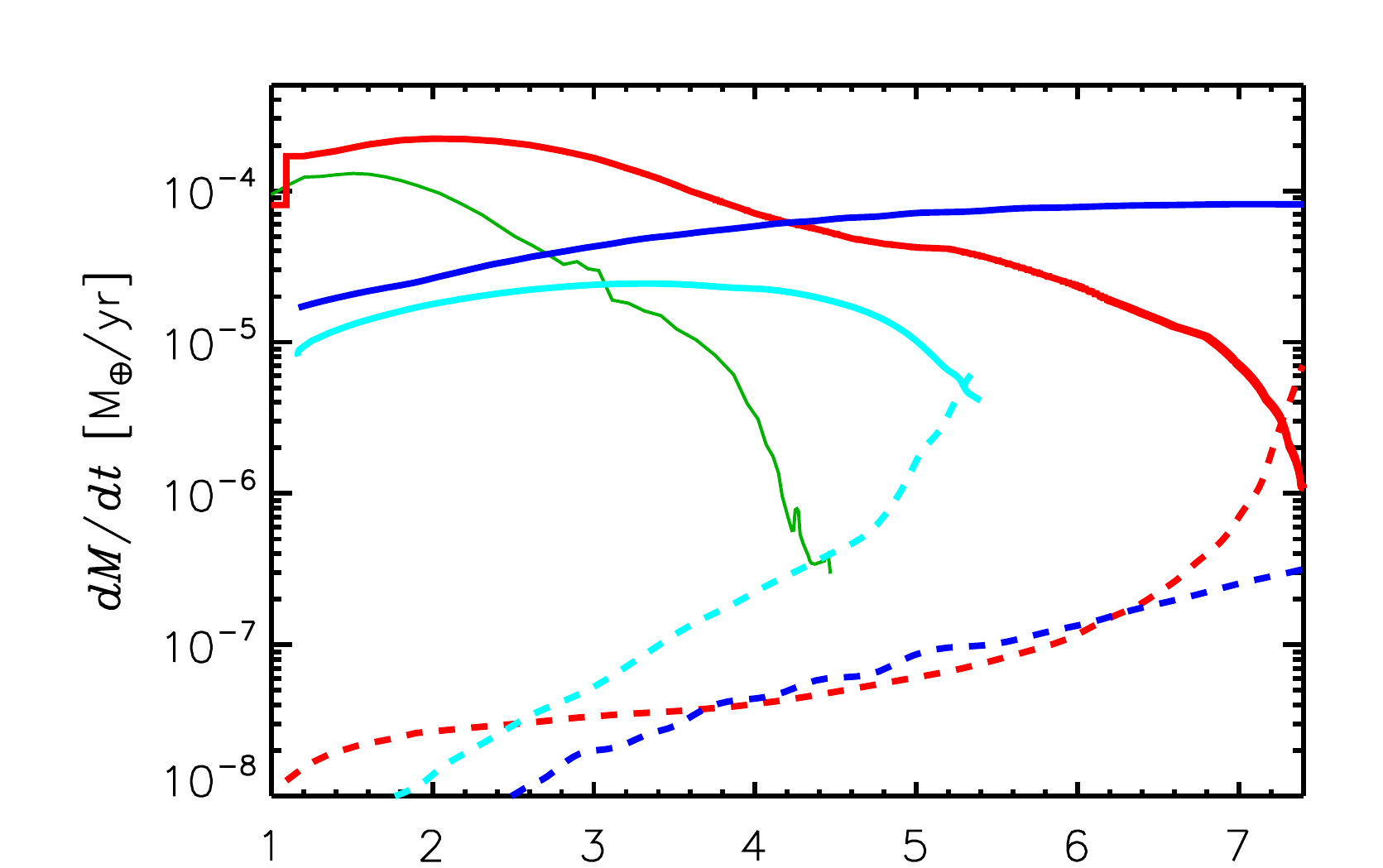}}
 \resizebox{\hsize}{!}{\includegraphics[angle=00]{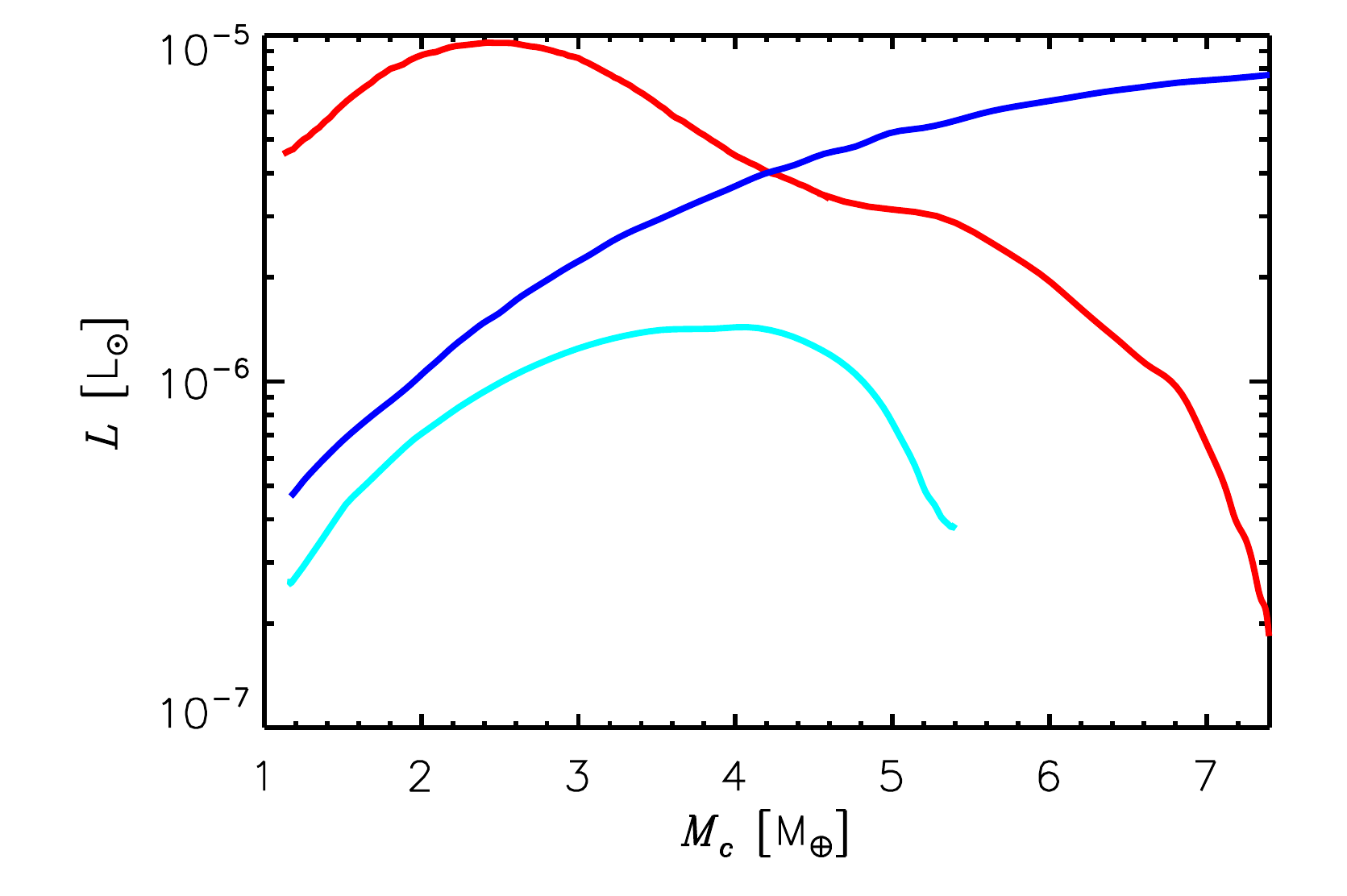}}
  \caption{%
               Top:
               Accretion rate of solids (red solid line) and of gas (red dashed line)
               as function of the core mass according to the Jupiter formation model
               presented herein. The blue and cyan lines show the results 
               for same quantities obtained, respectively, from models $\sigma$10 
               and $\sigma$6 of \citetalias{naor2010} (see also Figure~\ref{fig:mcvst}).
               The thin green solid line represents the accretion rate of solids for our
               calculation that does not account for the enhancement of the effective
               collision radius of planetesimals due to the presence of the envelope 
               around the core
               (see discussion in Section~\ref{sec:EPS} and Figure~\ref{fig:mc_bare}). 
               The accretion rates plotted here are averaged over time intervals
               of $10^{3}$--$10^{4}$ years.
               Bottom:
               Planet luminosity versus core mass for the cases corresponding to
               the models in the top panel (luminosity is not available for the bare
               core calculation).
               }
  \label{fig:naco}
\end{figure}

In absence of the envelope, there is a rapid drop of $\dot{M}_{c}$ for 
$M_{c}\gtrsim 2\,\mathrm{M}_{\oplus}$ (see top-right panel, see also Figure~\ref{fig:naco}),
which instead becomes a gradual reduction when gas drag in the envelope is at work 
(see the top-left panel). 
Comparing distributions in the top panels, the accretion rates of the 
$3\,\mathrm{M}_{\oplus}$ core without envelope are comparable to those 
of the $6\,\mathrm{M}_{\oplus}$ core with envelope. In fact, the cumulative 
distributions show that the total accretion rate of the $3\,\mathrm{M}_{\oplus}$ 
bare core is similar (a $50$\% difference) to that of the $6\,\mathrm{M}_{\oplus}$ 
core with envelope. 
The $4\,\mathrm{M}_{\oplus}$ bare core has accretion rates very similar to 
those of the $7\,\mathrm{M}_{\oplus}$ core with envelope for radii
$R\lesssim 1\,\mathrm{km}$ and $R\gtrsim 100\,\mathrm{km}$. 
The peak accretion occurs at $R\approx 50\,\mathrm{km}$ for both distributions, 
but it is less than $1/4$ as large in the bare core calculation.
The cumulative distributions in Figure~\ref{fig:dmcdt} also indicate that 
about $10$\%, or less, of the accreted mass is in bodies smaller than 
$100\,\mathrm{m}$ in radius and about $70$\% is in bodies larger than 
$10\,\mathrm{km}$. This last percentage varies somewhat,
from $\approx 60$\% for $M_{c}=1$--$2\,\mathrm{M}_{\oplus}$ to 
$\approx 80$\% for $M_{c}\gtrsim 6\,\mathrm{M}_{\oplus}$, possibly
due to the transfer of mass from smaller to larger planetesimals as
the swarm evolves (see Section~\ref{sec:EPS}).
As explained below, the fact that most of the solids mass passing through the
envelope is carried by large, $10\,\mathrm{km}$-size bodies may have 
important consequences for the accretion of gas and the ultimate formation
timescale. In fact, if most mass
was contained in bodies of much smaller size, more dust would be deposited
in the envelope and its opacity could be higher, inhibiting cooling and hence 
contraction.

A comparison among various calculations of solids and gas accretion rates,
versus core mass, is presented in the top panel of Figure~\ref{fig:naco}. The solid lines 
represent $\dot{M}_{c}$ and the dashed lines $\dot{M}_{e}$. The red lines
refer to the model discussed here. The blue and cyan lines show the results 
of models labelled as $\sigma$10 ($\sigma_{Z}=10\,\mathrm{g\,cm^{-2}}$ at
$5.2\,\mathrm{AU}$) and $\sigma$6 ($\sigma_{Z}=6\,\mathrm{g\,cm^{-2}}$
at $5.2\,\mathrm{AU}$) of \citetalias{naor2010}. 
In those calculations, the solids accretion rates were based on the three-body problem 
accretion calculations of \citet{greenzweig1992}, assuming a single-size planetesimals
of $100\,\mathrm{km}$ in radius. The condition $\dot{M}_{e}\approx \dot{M}_{c}$ occurred
at around the isolation mass given by Equation~(\ref{eq:Miso}),
respectively $11.5$ and $5.6\,\mathrm{M}_{\oplus}$, whereas here
the condition is realized prior to reaching that mass. However, in the
present calculations, the concept of isolation mass (at least as stated in 
Equation~(\ref{eq:Miso})) does not strictly apply because 
of interactions among planetesimals and the effects of drag forces 
(see discussion Section~\ref{sec:EPS}).

Figure~\ref{fig:naco} (top panel) also shows, as a thin green solid line, 
$\dot{M}_{c}$ for the case in which the enhancement of the cross-section 
for solids accretion due to the envelope is not included. 
Although the green and red lines have comparable
peaks ($\approx 1.3\times 10^{-4}$ and 
$\approx 2.2\times 10^{-4}\,\mathrm{M}_{\oplus}\,\mathrm{yr}^{-1}$,
respectively) and the maximum of $\dot{M}_{c}$ is reached
at similar values of the core mass ($M_{c}\approx 1.5$ and $2.1\,\mathrm{M}_{\oplus}$), 
the presence of the envelope already accounts for a factor of $10$ difference 
in $\dot{M}_{c}$ at $M_{c}=3.5\,\mathrm{M}_{\oplus}$ and a factor $100$ 
difference at $M_{c}=4.3\,\mathrm{M}_{\oplus}$.
It is not trivial to predict the conditions for which the maximum of $\dot{M}_{c}$ 
occurs. The accretion rate is mainly determined by the surface density 
of the planetesimal swarm, which is reduced by shepherding. 
The effectiveness of shepherding is a function of damping, which
is due to two processes: collisions and gas drag. The rate of collisional 
damping depends on both the surface density of the swarm and the 
size distribution of the bodies, while the gas drag term depends on the size 
of an individual body and on the local gas density.
However, the capture radius for accretion of 
$10\,\mathrm{km}$-radius (and larger) planetesimals, which contribute 
most of the accreted mass (see Figure~\ref{fig:dmcdt}), begins to be affected 
by the gaseous envelope at core masses between $1$ and $2\,\mathrm{M}_{\oplus}$.
Therefore, somewhat different values of $M_{c}$ at which the accretion rates peak 
(and the maximum values) in the calculations with and without envelope
are expected.

The luminosity of the planet as a function of the core mass is shown in the bottom panel 
of Figure~\ref{fig:naco} for the same models as in the top panel, except for the calculation
without the envelope. The release of potential energy by both envelope contraction and 
solids accretion represents the source of planet luminosity. However, during these early
phases of growth, most of the luminosity is provided by accretion of solids. In fact, 
the luminosity peaks in the figure closely follow the peaks in $\dot{M}_{c}$.
In the current calculation, 
the maximum of the luminosity occurs $\approx 2000$ years later than the maximum
of the solids accretion.

\begin{figure*}
\centering%
 \resizebox{\hsize}{!}{\includegraphics[angle=00]{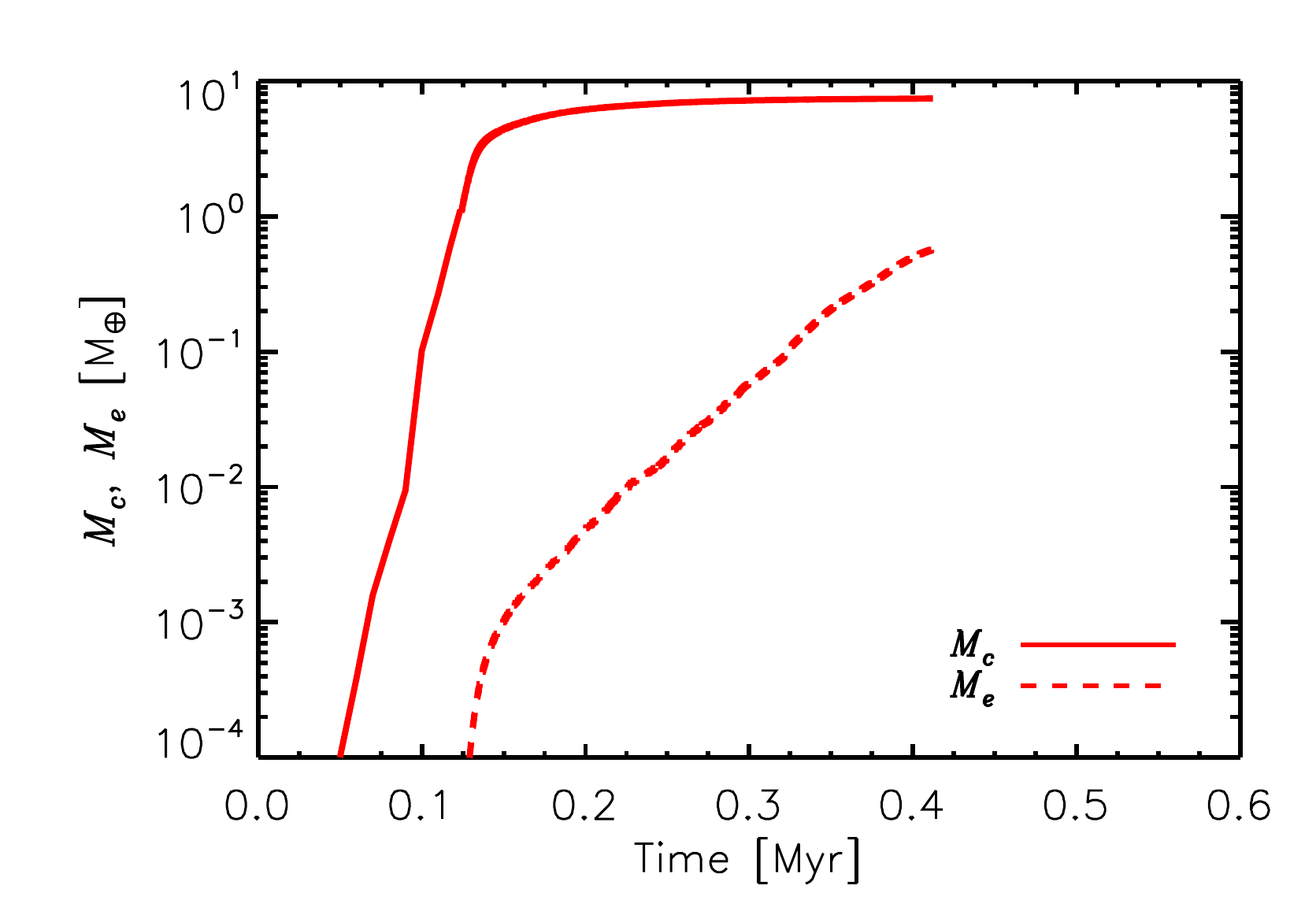}%
                                  \includegraphics[angle=00]{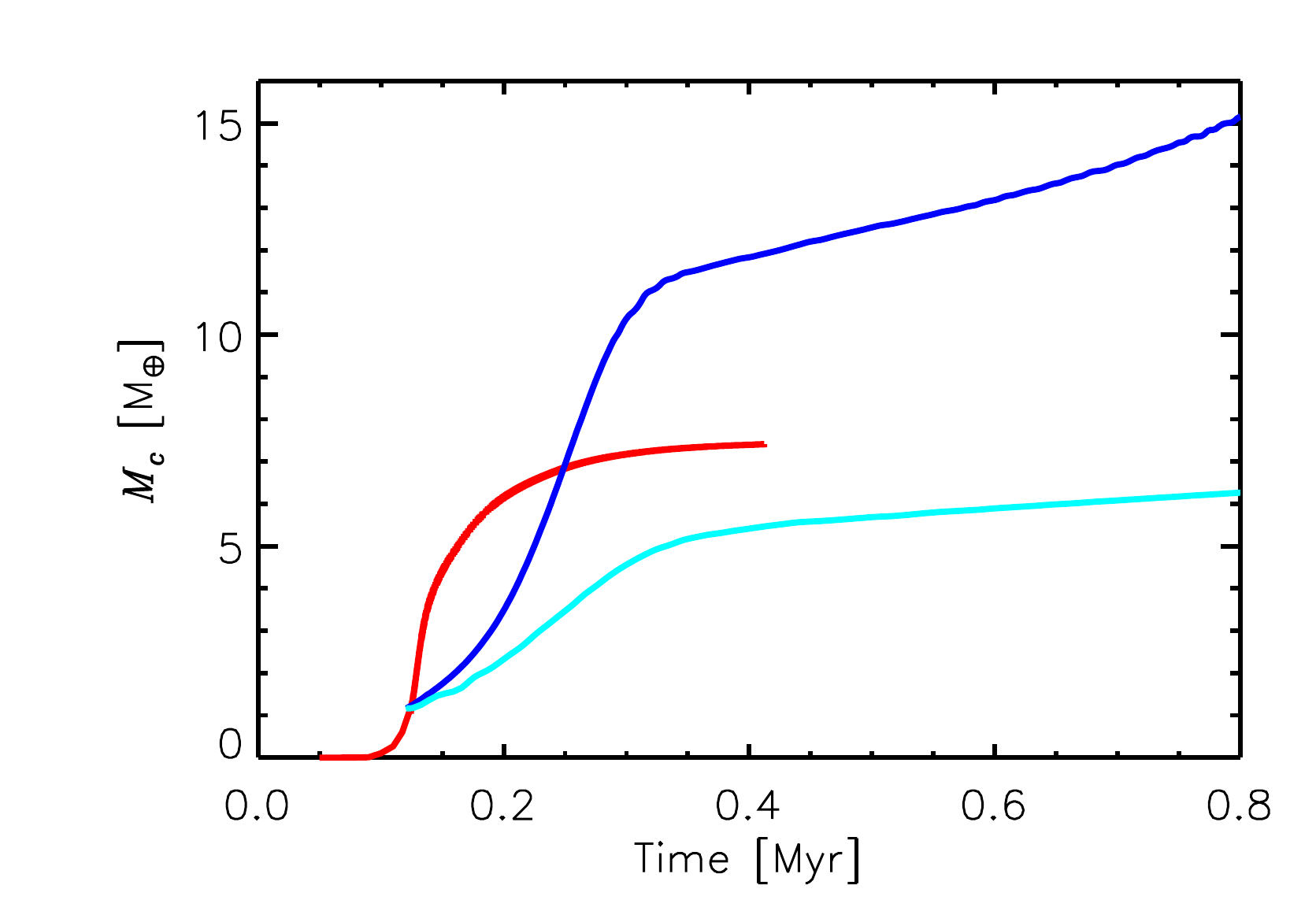}}
  \caption{%
               Left:
               Core ($M_{c}$) and envelope mass ($M_{e}$) versus time
               in the current calculation.
               Right:
               Core mass growth for various models. The red line represents
               the current model (the same as in the left panel).
               The blue and cyan lines show the evolution in models $\sigma$10 
               and $\sigma$6 of \citetalias{naor2010} (see also Figure~\ref{fig:naco}),
               based on the solids accretion rates of \citet{greenzweig1992} and a 
               single-size distribution with planetesimals of $100\,\mathrm{km}$ in radius. 
               These last two models are time shifted so that
               $M_{c}\approx 1.1\,\mathrm{M}_{\oplus}$ at approximately the same 
               time for all curves. 
               }
  \label{fig:mcvst}
\end{figure*}
The temporal evolution of both core and envelope mass may differ quite substantially 
from that of previous calculations that do not simulate the evolution of the 
planetesimal swarm, as illustrated in Figure~\ref{fig:mcvst}. 
For example, despite having the same solids surface density 
at $5.2\,\mathrm{AU}$, $M_{c}$ in model $\sigma$10 of \citetalias{naor2010}
grows much more rapidly, as $\dot{M}_{c}$ starts to decrease only when
$M_{c}\gtrsim 8\,\mathrm{M}_{\oplus}$, reaching 
$\approx 12\,\mathrm{M}_{\oplus}$ after $0.4\,\mathrm{Myr}$
(using the synchronization of Figure~\ref{fig:mcvst}, left panel).
At this point of the evolution,
the $\approx 1.6$ factor difference in core mass is reflected by a similar 
fractional difference in the envelope mass, $\approx 1\,\mathrm{M}_{\oplus}$ 
against $0.57\,\mathrm{M}_{\oplus}$.
There are some similarities to the $\sigma$6 model of \citetalias{naor2010} 
in that the core mass differs by about $2\,\mathrm{M}_{\oplus}$ 
(a $\approx 30$\% difference) after $0.4\,\mathrm{Myr}$ and the envelope 
mass is only somewhat smaller than in the previous calculation 
($M_{e}\approx 0.46\,\mathrm{M}_{\oplus}$). The larger $\dot{M}_{e}$
of the $\sigma$6 model for equal core masses 
(and $M_{c}\gtrsim 2.5\,\mathrm{M}_{\oplus}$, see Figure~\ref{fig:naco}), 
is a result of the lower solids accretion, which favors the contraction of 
the envelope on shorter timescales.

\begin{figure}
\centering%
 \resizebox{\hsize}{!}{\includegraphics[angle=00]{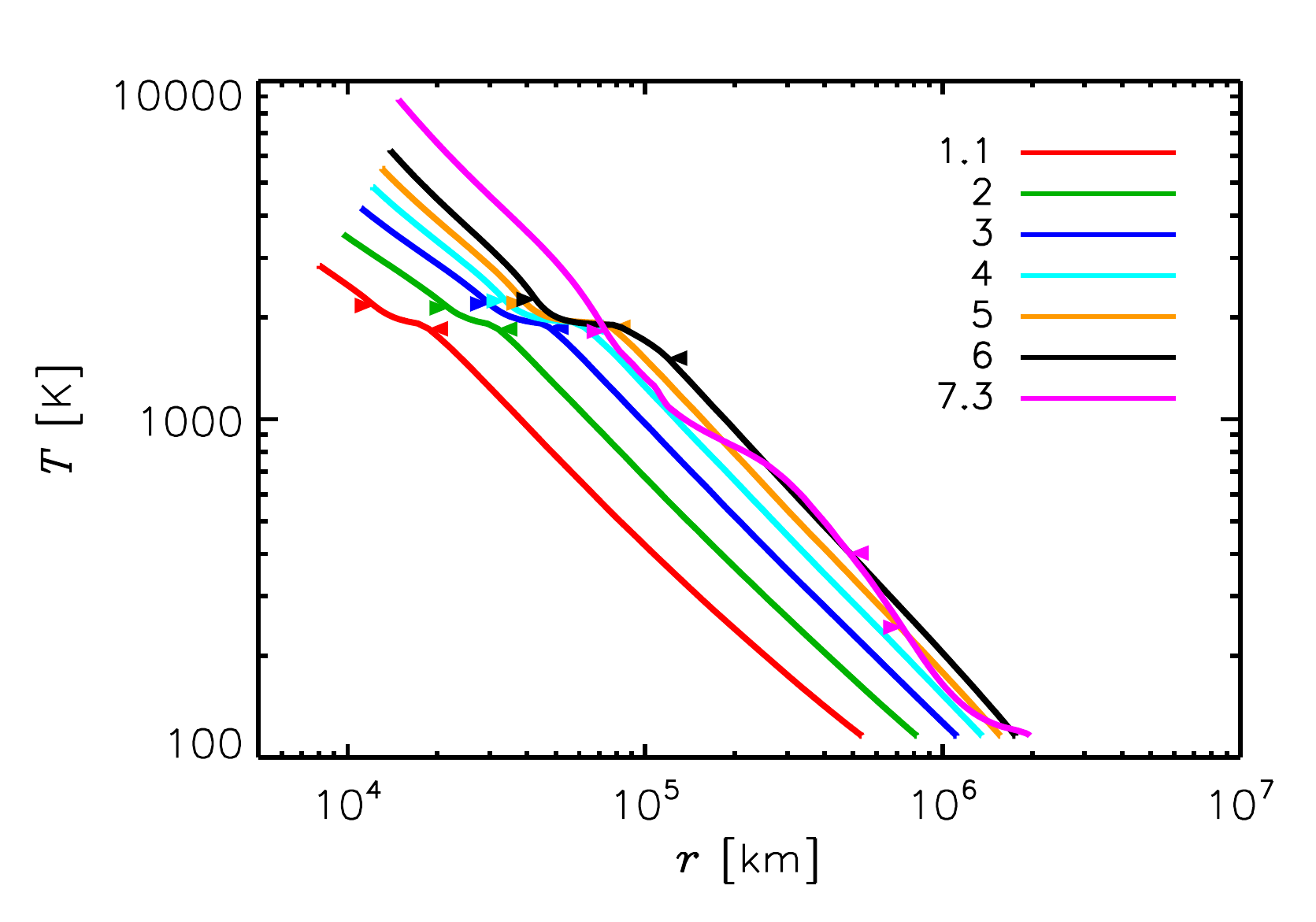}}
 \resizebox{\hsize}{!}{\includegraphics[angle=00]{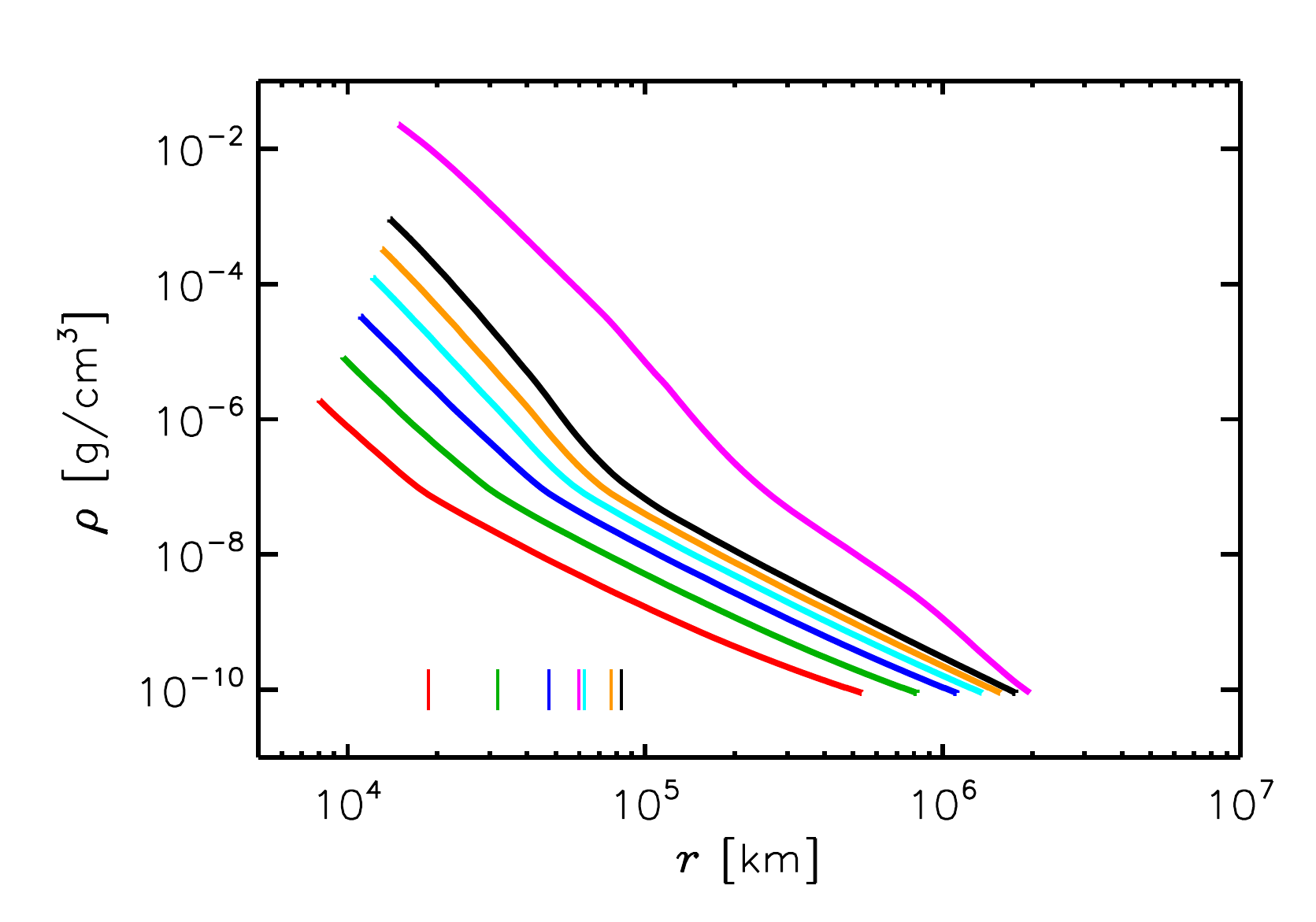}}
  \caption{%
               Temperature (top panel) and density (bottom panel) profiles
               of the planet's envelope for various values of the core mass, 
               as indicated in the legend of the top panel in units of $\mathrm{M}_{\oplus}$.
               The temperature decrease in the outer  layers between 
               the envelope models at $M_{c}=6$ and $7.3\,\mathrm{M}_{\oplus}$
               is associated with the decrease in the core accretion rate
               (see top panel of Figure~\ref{fig:naco}). 
               The arrowheads (top) and the vertical line marks (bottom) indicate,
               respectively, the boundaries between convective and radiative layers
               and the radii where molecular hydrogen begins to dissociate.
               }
  \label{fig:prcm1}
\end{figure}
\begin{figure}
\centering%
 \resizebox{\hsize}{!}{\includegraphics[angle=00]{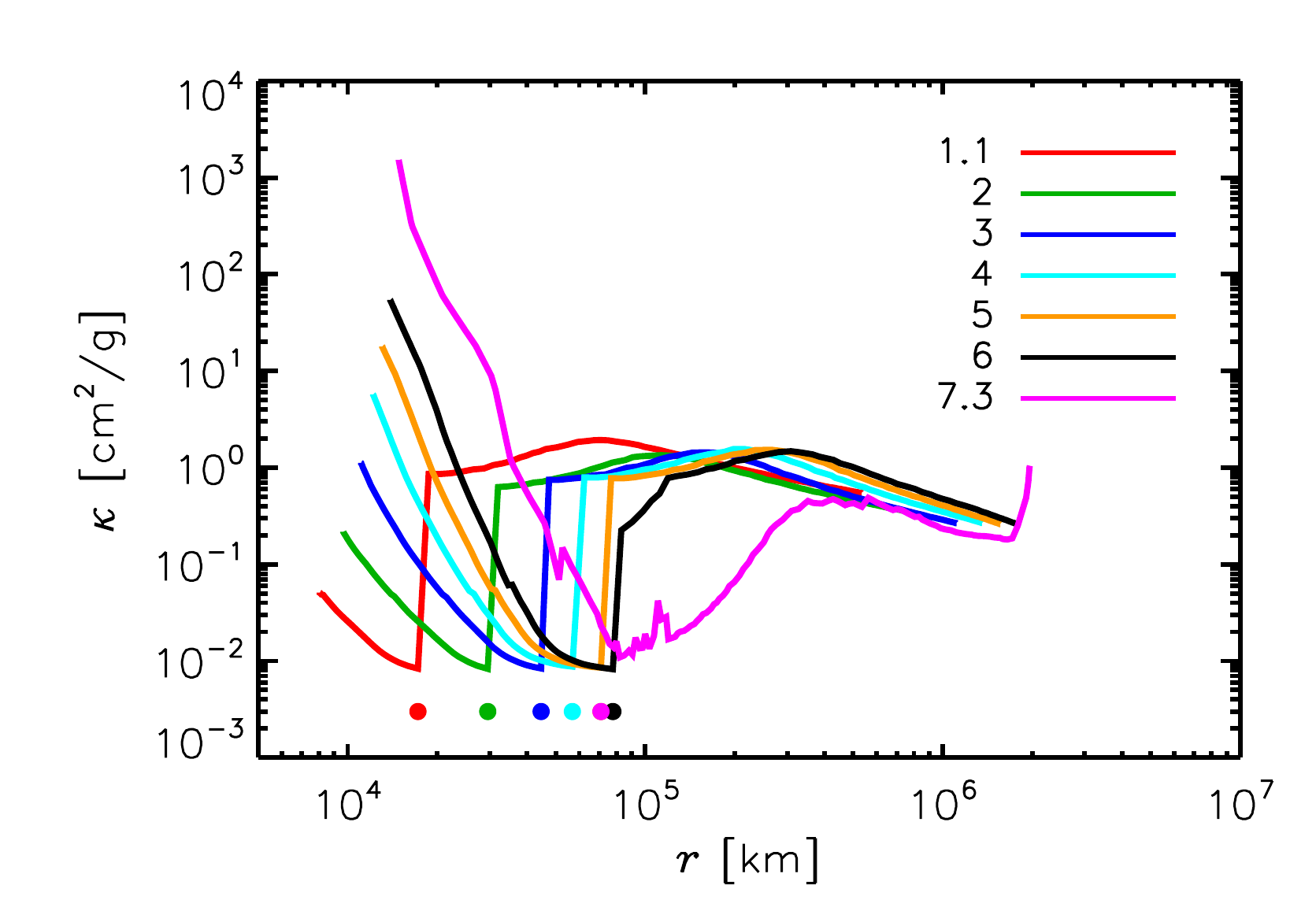}}
  \caption{%
               Opacity profiles of the envelope for the planet at different values
               of the core mass, as
               indicated in the legend in units of $\mathrm{M}_{\oplus}$. 
               The opacity is dominated by dust contributions
               at temperatures $T\lesssim 1800\,\mathrm{K}$.
               In these envelope layers, the opacity is smaller than
               typical interstellar dust opacity by factors between $\sim 10$
               and $\sim 1000$.
               The symbols underneath the curves indicate the radii corresponding
               to the temperature at which
               dust grains are assumed to evaporate. Gas opacity dominates
               the opacity coefficient inside those radii.
               }
  \label{fig:prcm2}
\end{figure}

In Figure~\ref{fig:prcm1}, we plot the temperature (top) and density (bottom)
structure of the envelope for different values of the core mass.
The temperature plateau at $\sim 2000\,\mathrm{K}$ is the result of low 
molecular opacity (see the drop of $\kappa$ in Figure~\ref{fig:prcm2}) 
and low density ($\rho\lesssim 10^{-6}\,\mathrm{g\,cm^{-3}}$) 
at temperatures just above the evaporation temperature 
of the dust. Dissociation of molecular hydrogen may also contribute somewhat,
as indicated by the line marks beneath the curves in the bottom panel.
The vertical line segments correspond to the radii at which significant 
dissociation begins: the density ratio of atomic to molecular hydrogen
is $\gtrsim 0.01$ to the left of the line marks.
This temperature plateau is absent at $M_{c}=7.3\,\mathrm{M}_{\oplus}$
mainly because of the higher gas densities in that region
($\rho\sim 10^{-4}\,\mathrm{g\,cm^{-3}}$), resulting in higher
molecular opacities and a higher optical thickness of that envelope layer.
In fact, at this core mass, Figure~\ref{fig:prcm2} indicates that there is no 
opacity drop around $2000\,\mathrm{K}$.
In addition, at those densities, significant dissociation of molecular hydrogen 
begins at higher temperatures,
as shown by the corresponding line segment in the bottom panel.
The change in temperature gradient around $700\,\mathrm{K}$ in the 
$M_{c}=7.3\,\mathrm{M}_{\oplus}$ case, at distances between
$1.5$ and $3 \times 10^{5}\,\mathrm{km}$,
appears associated with a low opacity around those locations
(see Figure~\ref{fig:prcm2}). 
The opacities at distances of $\sim10^{5}\,\mathrm{km}$ 
($400\lesssim T \lesssim 1500\,\mathrm{K}$) come primarily 
from the dust, and the drop visible between core masses of 
$5$ and $7.3\,\mathrm{M}_{\oplus}$ likely occurs partly because 
of the decline in $\dot{M}_{c}$, which causes less solid material 
to be delivered in those envelope layers. 
Moreover, because of the increase in envelope mass, planetesimals 
are ablated higher up in the envelope and so grains have more time 
to coagulate before they settle to those deeper layers, thus reducing 
the opacity.

The envelope structure surrounding cores with mass
$1.1\lesssim M_{c}\lesssim6.3\,\mathrm{M}_{\oplus}$ are composed
of two convective shells separated by a radiative shell. 
The arrowhead symbols in the top panel of Figure~\ref{fig:prcm1} mark 
the radial boundaries between convective and radiative layers, pointing 
in the direction in which the radiative zones extend.
The outer radius of the interior convective zone increases as the core and
envelope mass grow. The inner radius of the exterior convection zone
also expands outward.
Nonetheless, when $M_{c}\approx 6.3\,\mathrm{M}_{\oplus}$
($M_{e}\approx 6\times 10^{-3}\,\mathrm{M}_{\oplus}$), the outer 
convection shell still occupies over $99.9$\% of the envelope volume
(corresponding to $30$\% of the mass of the envelope).
As $M_{c}$ and $M_{e}$ increase further, the outer layers become
radiative. 
At $M_{c}\approx 7.3\,\mathrm{M}_{\oplus}$, there are two convective
and two radiative zones of which the outermost (radiative) zone occupies
$96$\% of the envelope volume (comprising $10$\% of the mass).

Simple arguments based on dissipation of kinetic energy via gas drag
suggest that the path of a body traveling through gas is significantly 
affected once it encounters a mass of gas about equal to its own mass
(see Section~\ref{sec:CRC}). Thus, assuming a traveling distance in the
envelope on the order of the envelope radius $R_{p}$ 
(i.e., $R_{\mathrm{capt}}\ll R_{p}$), captured planetesimals
have radii $R\sim (\rho/\rho_{s}) R_{p}$, where $\rho$ is an average envelope
density and $\rho_{s}$ the density of the body. 
For small objects, the traveling distance is $R_{p}-R_{\mathrm{capt}}$ and
one should iterate the relation $R\sim (\rho/\rho_{s}) (R_{p}-R_{\mathrm{capt}})$.
Gas densities corresponding 
to the envelope around a $7.3\,\mathrm{M}_{\oplus}$ core imply that
$\sim 100\,\mathrm{km}$ size planetesimals 
may be captured at a distance of $\sim 8\times 10^{4}\,\mathrm{km}$.
Planetesimals of $\sim 10$  and $\sim 1\,\mathrm{km}$ in size may be captured 
at about $1.5$ and $2.5$ times that distance, respectively. 
These numbers agree within factors of order unity with the capture 
radii shown in Figure~\ref{fig:rcapt}, but in reality 
the situation is more complex. In fact, planetesimal capture results from the
combined effects of kinetic energy dissipation and mass loss due to ablation.
As a body sheds mass, dissipation of energy by drag is facilitated.

For a given material, the mass loss of a body via ablation is proportional
to its surface \citep[see, e.g.,][]{podolak1988}. 
For an amount of solids $\Delta M$ traveling through an envelope layer
during a time interval $\Delta t$,
the mass ablated, $\Delta M_{\mathrm{ab}}$, is proportional to the total
surface area exposed by the solids. 
Therefore, the ratio $\Delta M_{\mathrm{ab}}/\Delta M$
is proportional to the total surface to volume ratio, i.e., to $1/R$ assuming a single
size distribution of bodies with radius $R$.
This implies that, for equal masses of solids accreted in planetesimals of different
sizes, smaller planetesimals are expected to shed more mass, and hence release 
more dust in the envelope, than larger planetesimals are.

Figure~\ref{fig:prcm2} shows the opacity in the envelope for increasing
core masses. In the outer envelope, the opacity is entirely due to dust. 
At temperatures around $1800\,\mathrm{K}$, silicates are assumed to 
evaporate and opacity is dominated by gas above that temperature.
The solid circles in the figure indicate the approximate radii
in the envelopes at which dust grains evaporate and molecules
dominate the opacity coefficient \citep{freedman2008}. 
The accretion of gas is dictated by the contraction of the envelope,
which is regulated by the ability of the outer layers to radiate energy
away, which in turn depends on the opacity of these layers. 
The figure indicates that $\kappa$ is between $\sim 0.01$ and 
$\sim 1\,\mathrm{cm^{2}\,g^{-1}}$, one to three orders of
magnitude smaller than the typical interstellar dust opacity.
The minimum dust opacities are higher than those in the
models of \citetalias{naor2010}. The procedures for the opacity
calculation are the same, but in the current study we include much
smaller planetesimals. The difference may be caused in part by ablation 
of small planetesimals, although the envelope structure 
and solids accretion rates are different and therefore other effects may be 
important as well.

The opacity of the outer envelope layers is determined by the accretion rate
of the gas, rather than by that of the solids. 
If the accretion flow delivers small grains to these layers on a timescale
shorter than that required by the grains to coagulate and settle, the
opacity is affected. Therefore, as $\dot{M}_{e}$ begins to rise significantly, 
a gradual increase of $\kappa$ should be expected at the outer boundary 
over time. Indeed, for $M_{c}>6\,\mathrm{M}_{\oplus}$, we observe
that the opacity of the outermost layers steadily increases, with some
fluctuations corresponding with variations of the gas accretion rate.

The top panel of Figure~\ref{fig:naco} shows that gas accretion starts
to increase substantially only after solids accretion drops well below its
maximum value. 
However, as argued above, the opacity is not only affected by the input 
rate of solids, $\dot{M}_{c}$, but also by the size of the bodies carrying 
the accreted mass.
The accretion rates peak at planetesimals' radii between $30$ and $50$ 
kilometers. If the peak moved to smaller radii, a larger quantity of dust would be 
released in the outer envelope layers, increasing the opacity, delaying contraction, 
and reducing or possibly inhibiting the accretion of gas.

\section{Summary and conclusions}
\label{sec:summary}

We have modeled the growth of Jupiter's core at $5.2\,\mathrm{AU}$ from 
the proto-sun within a solar composition disk having a planetesimal surface 
mass density $\sigma_{Z} = 10 (5.2\,\mathrm{AU}/a)\,\mathrm{g\,cm^{-2}}$, 
where $a$ is the heliocentric distance. 
Our simulations follow the evolution of solid bodies (see Figures~\ref{fig:maps}
and \ref{fig:ei_maps}) 
using the multi-zone accretion 
code developed by \citet{weidenschilling1997} and \citet{weidenschilling2011}, 
and modified 
as described in Section~\ref{sec:PAC}. This code accounts for gravitational 
interactions and gas drag, and assumes that physical collisions between
solid bodies lead to accretion, or with sufficiently high impact velocities, 
to fragmentation.  
The planetary embryo begins as a Ceres-mass solid body
(see Figure~\ref{fig:mcvst}), but once it grows to 
a mass of $1.1\,\mathrm{M}_{\oplus}$ it begins to accumulate a gaseous 
envelope whose quasi-hydrostatic structure and accretion of gas are calculated 
by means of the formalism described in Sections~\ref{sec:ESC} through \ref{sec:CM}.

Relatively small fractions of the total mass of planetesimals are contained
in bodies smaller than $1\,\mathrm{km}$ or larger than $100\,\mathrm{km}$ 
in radius. In fact, typically $< 9$\% of the mass of the swarm is in planetesimals 
smaller than $1$ kilometer in radius, $\lesssim 25$\% is in planetesimals with 
radii between $1$ and $10$ kilometers, and $\lesssim 7$\% is accounted for 
by bodies with radii larger than $100$ kilometers. As the swarm evolves, 
mass is preferentially transferred to larger bodies. 
In particular, at the time when $M_{c}\approx 7\,\mathrm{M}_{\oplus}$, 
about $80$\% of the mass is in the form of planetesimals with radii
$10\lesssim R\lesssim 130\,\mathrm{km}$, and about $60$\% is accounted 
for by planetesimals whose radii are between $\sim 20$ 
and $\sim 80\,\mathrm{km}$ (see Figure~\ref{fig:cum_mass}).

The core's growth rate initially accelerates rapidly via runaway accretion, 
but it eventually declines as accretion and shepherding combine to substantially 
depress the surface density of planetesimals in orbits surrounding the planet
(see Figure~\ref{fig:maps}). The core growth rate peaks at 
$2.2\times 10^{-4}\,\mathrm{M}_{\oplus}\,\mathrm{yr}^{-1}$ when 
$M_{c}\approx 2.1\,\mathrm{M}_{\oplus}$ (see Figure~\ref{fig:naco}). 
Although the mass of the envelope remains small 
($M_{e}\lesssim 10^{-3}\,\mathrm{M}_{\oplus}$) until that of the core 
exceeds about $5\,\mathrm{M}_{\oplus}$, the envelope is large in volume, 
and is sufficiently dense to substantially increase the accretion rate of the core 
(by increasing the planet's capture cross-section for small planetesimals) once 
the planet's mass reaches $2\,\mathrm{M}_{\oplus}$ (see Figures~\ref{fig:rcapt}
and \ref{fig:dmcdt}). 
The gas accretion rate gradually increases, and becomes equal to the accretion 
rate of solids after about $4\times 10^{5}$ years. At this time, the planet's core 
mass is $M_{c}\approx 7.3\,\mathrm{M}_{\oplus}$ and its envelope mass is 
$M_{e}\approx 0.15\,\mathrm{M}_{\oplus}$ (see Figures~\ref{fig:naco}
and \ref{fig:mcvst}). 
A planet lacking an envelope would only grow to $4.4\,\mathrm{M}_{\oplus}$ 
at this time in the same disk (see Figure~\ref{fig:mc_bare}), and its growth 
rate at this point would be very small, 
$\dot{M}_{c}\approx 3\times 10^{-7}\,\mathrm{M}_{\oplus}\,\mathrm{yr}^{-1}$
(see Figure~\ref{fig:naco}), as a consequence of gap clearing in the planetesimals' disk. 
In this case, because the shepherding effect depends on dissipation by collisions 
(mostly) and gas drag, we find that the mass limit is around $40$\% of the standard 
isolation mass (see Equation~(\ref{eq:Miso})).

We compare the results with previous models, which rely on estimates of solids 
accretion rates based on the three-body problem accretion calculations and
planetesimals of a single size, and show that there are substantial differences
(see Figure~\ref{fig:naco}).
One such calculation \citep{naor2010}, with the same $\sigma_{Z}$ and 
the same envelope physics as the present simulation, gives the result, 
at the end of Phase~1 ($t\approx 0.45\,\mathrm{Myr}$), of 
$M_{c}=11.5\,\mathrm{M}_{\oplus}$ and $M_{e}=0.67\,\mathrm{M}_{\oplus}$.

For a single size distribution of planetesimals with radius $R$, the mass ablated
in an envelope layer is a fraction of the accreted mass (passing through that
layer) proportional to $1/R$ (see discussion in Section~\ref{sec:results}).
Therefore, smaller accreted bodies shed larger amounts of dust,
possibly raising the envelope opacity \citep{naor2008} and slowing envelope contraction.
The fact that most of the accreted solid mass is in the form of relatively
large planetesimals may favor the early growth of the gaseous envelope.
Our results demonstrate the influence of a low-mass but voluminous planetary 
envelope on planetesimal accretion, and imply that Jupiter's core could have 
accumulated at the planet's current location in a protoplanetary disk whose surface 
mass density is only a few times as large as that of a classical minimum mass solar 
nebula.

Although the present study takes into account many physical processes 
relevant to the formation of a giant planet's core in a solar-type nebula, 
a number of effects are neglected.
Among these are gas- and planetesimal-driven migration, trapping of bodies 
into mean motion resonances with the core, and the growth from the swarm
of competing embryos.
Inside $\sim 5\,\mathrm{AU}$, gas-driven migration is thought to push inward 
cores less massive than about $10\,\mathrm{M}_{\oplus}$ \citep{baruteau2014}
at a rate that appears very sensitive to the local (and time-varying) 
thermodynamical properties of the nebula gas.
Planetesimal-driven migration in a disk of solids whose dynamics is dominated 
by Keplerian shear can potentially operate inward as well as outward 
\citep[e.g.,][and references therein]{kirsh2009,levison2010,capobianco2011},
depending on the details of both the planetesimal and the gas disks.
Capture of bodies into mean motion resonances with the core may push it toward 
the sun \citep[e.g.,][]{levison2010}, if the core is allowed to migrate. However, as the
number density of bodies in and around the resonance region increases, collisions 
between planetesimals can become effective at altering their residence time in the 
resonance \citep{weidenschilling1985}.
As already mentioned in Section~\ref{sec:EPS}, our calculations do allow for 
the possibility of oligarchic growth of multiple cores in the swarm of planetesimals 
and had we not started with a Ceres-sized seed body, multiple cores might have 
emerged. 
However, the seed body (which is only a factor of several larger than the largest 
bodies in the initial swarm) grows rapidly enough to inhibit potential competitors 
due to its perturbations.
Clearly, a comprehensive calculation should attempt to include 
all of these (and other) effects, along with those included in this study.

\section*{Acknowledgments}

This project was funded by NASA Outer Planets Research Program Grant 202844.02.02.01.75.
We are grateful to Olenka Hubickyj for helpful discussions and useful feedback on this work,
and to two anonymous referees whose comments and suggestions helped improve this paper. 
G.D.\ thanks Los Alamos National Laboratory for its hospitality.
Resources supporting this study were provided by the NASA High-End
Computing (HEC) Program through the NASA Advanced Supercomputing
(NAS) Division at Ames Research Center and the NASA Center for Climate 
Simulation (NCCS) at Goddard Space Flight Center.








\end{document}